\documentclass[%
 reprint,
bibnotes,
 amsmath,amssymb,
prb,
longbibliography
]{revtex4-2} 
\usepackage{graphicx}
\usepackage{dcolumn}
\usepackage{bm}

\usepackage{hyperref}

\begin{document}

\preprint{APS/123-QED}

\title{Trapped-mode excitation in all-dielectric metamaterials with loss and gain}

\author{Anton~V.~Hlushchenko$^{1,2}$}
\author{Denis~V.~Novitsky$^{1}$}
\author{Vladimir~R.~Tuz$^{1,3}$}
\email{tvr@jlu.edu.cn}
\affiliation{$^1$State Key Laboratory of Integrated Optoelectronics, College of Electronic Science and Engineering, International Center of Future Science, Jilin University, 2699 Qianjin Street, Changchun, 130012, China}
\affiliation{$^2$National Science Center `Kharkiv Institute of Physics and Technology' of National Academy of Sciences of Ukraine, 1, Akademicheskaya Street, Kharkiv 61108, Ukraine}
\affiliation{$^3$School of Radiophysics, Biomedical Electronics and Computer Systems, V.~N.~Karazin Kharkiv National University, 4, Svobody Square, Kharkiv 61022, Ukraine}

\date{\today}

\begin{abstract}
Non-Hermitian photonics based on combining loss and gain media within a single optical system provides a number of approaches to control and generate the flow of light. In this paper, we show that by introducing non-Hermitian perturbation into the system with loss and gain constituents, the high-quality resonances known as trapped modes can be excited without the need to change the symmetry of the unit cell geometry. To demonstrate this idea, we consider a widely used all-dielectric planar metamaterial whose unit cell consists of a pair of rectangular nanoantennas made of ordinal (with loss) and doped (with gain) silicon. Since the quality factor of the trapped-mode resonance can be controlled by changing both spatial symmetry and non-Hermiticity, varying loss and gain allows us to compensate for the influence of asymmetry and restore the quality factor of the localized mode. The results obtained suggest new ways to achieve high-quality resonances in non-Hermitian metamaterials promising for many practical applications in nanophotonics.
\end{abstract}

\maketitle

\section{Introduction}

One of the key aims of active nanophotonics is to develop advanced nanodevices that provide the most efficient interaction of light with nanostructured matter for lasing and optical switching operations \cite{krasnok2020active}. Artificial nanostructures with superior functionalities as compared to natural materials make it possible to control both the propagation of light and optical dynamics by changing the sign and profile of complex permittivity, as well as the ratio between its real and imaginary parts. Active nanophotonics is a platform for implementing the concepts of non-Hermitian physics and, in particular, the ideas of $\mathcal{PT}$ symmetry and exceptional points \cite{ruter2010observation, ozdemir2019parity}. There is a number of applications of these ideas to non-Hermitian structures consisting of elements with gain and loss \cite{kang2016chiral, alaee2018optical, dong2020loss, song2021plasmonic, wang2021active, yu2021dielectric, dong2022nanoscale}. Coupling these elements (cavities, waveguides, resonators, etc.) implies interaction between the modes of the system \cite{hlushchenko2021multimode, hlushchenko2021multimode2}. The interaction of modes in coupled non-Hermitian systems can lead to appearance of exceptional points where the modes become degenerate or $\mathcal{PT}$ symmetry gets broken \cite{ozdemir2019parity}.

On the other hand, the rapid development of nanotechnologies \cite{zheludev_NatureMat_2012, Koenderink_Scince_2015, Neshev_NatLight_2018} stimulates the scientific community to introduce new mechanisms for controlling optical processes associated with the interaction of light with artificial media called metamaterials. According to the substances used for their production, metamaterials can be divided into two major classes -- metallic and all-dielectric ones. To date, all-dielectric nanostructures have gained great popularity due to the fact that in the infrared and visible parts of the spectrum, they demonstrate significantly lower material losses compared to their metallic (plasmonic) counterparts. Moreover, they are compatible with the well-established CMOS technology promising for realizing many nanophotonic devices \cite{jahani2016all}. Although today the practical possibilities for constructing metamaterials have expanded significantly, nevertheless, controlling their response in a dynamic way may be problematic. It is often necessary to change the geometry of the system which is difficult to implement in practice. Therefore, new mechanisms are needed for controlling optical response, for example, by using structures with active elements containing gain media. The development of various technologies for incorporating active constituents in optical metamaterials resulted in the discovery of fundamentally new mechanisms for electromagnetic wave interaction with them \cite{Gu_NatureCom_2012, Ma_NatureNano_2019, Tong_AdvFuncMat_2019, Shaltout_Scince_2019}. Changing the properties of active elements is possible due to external manipulations on demand and is not limited to the specifics of the metastructure production process \cite{Rahmani_AdvFuncMat_2017, Shaltout_Scince_2019}. Dynamic control over the light propagation by tuning active elements of metamaterials has many practical applications, such as loss compensation \cite{Xiao_Nature_2010, Amooghorban_PRL_2013, ghoshroy2020loss},  lasing \cite{hess2012active, droulias2017fundamentals, deka2021nanolaser}, nonlinear optical operations \cite{tuz_PhysRevB_2010, tuz_JOpt_2012, Tuz_JOSAB_2014}, thermal radiation control \cite{dyachenko2016controlling}, interferometry \cite{dabidian2016experimental}, holography \cite{li2017electromagnetic, malek2017strain}, etc.

Among many configurations of all-dielectric metamaterials, we are interested here in two-dimensional flat structures (metasurfaces) that support the so-called trapped modes \cite{khardikov2012giant,zhang2013near,Khardikov2016,tuz2018high} (recently, such modes are also referred to as the phenomenon of bound states in the continuum (BICs) \cite{Hsu2016BIC, Azzam2020AOM, Sadreev2021rpp}). In metamaterials, these modes are related to purely real eigenstates existing in idealized lossless infinitely expanded structures whose translation unit cells possess specific spatial symmetry. To excite the corresponding eigenstate by the field of the incident radiation, a particular perturbation should be introduced into the unit cells which breaks their symmetry \cite{fedotov2007sharp}. The use of specific irradiation conditions (oblique incidence, near-field sources, etc.) is another possible way to realize excitation of such modes in metamaterials \cite{Tian_ACSPhoton_2014, Fan_Optica_2019, van2021unveiling}.

Hereinafter, as a basic example, we study a planar metamaterial with the unit cell consisting of two dielectric rectangular bars -- coupled dielectric nanoantennas (elements that are much smaller than the wavelength). This type of metamaterial has been studied earlier as a system supporting high-quality resonances \cite{khardikov2012giant, zhang2013near, Khardikov2016, Ndao_Nanophotonics_2020, gorkunov2021bound, kuznetsov2021transparent}. In the elementary unit cell (meta-atom) of such a system, the antiphased oscillations of displacement currents excited in the nanoantennas by incident light arise from the trapped mode when particular asymmetry is introduced into the size or position of the structural elements (bars). Thus, the degree of asymmetry determines the strength of interaction of the mode with the external incident field and, hence, the quality factor of the corresponding resonance.

When the unit cell of a metamaterial is composed of several particles, its electromagnetic properties are defined by modes supported by the unit cell as a whole \cite{khardikov2012giant, zhang2013near, tuz2018high, kang2016chiral, yu2021dielectric}. These modes arise as a result of the electromagnetic coupling between the modes of individual particles forming the unit cell. In particular, for a pair of coupled nanoantennas in the unit cell, there is a pair of modes supported by each nanoantenna which coalesce at the exceptional point \cite{yu2021dielectric}. On the other hand, the modes of the unit cell as a whole appear as an infinite set of separate modes distant from each other in the spectrum. Among such modes there are those that have purely real eigenfrequencies for a lossless (idealized) metamaterial. They belong to the class of trapped modes that are of great practical interest for metamaterial physics and applications.

In this paper, we study the non-Hermitian effects such as $\mathcal{PT}$-symmetry breaking \cite{Novitsky2021PRB} to implement active control over the trapped mode by utilizing loss and gain. In the lossless symmetric case (identical elements of the unit cell), this mode cannot be excited and has an infinite quality factor. In the presence of asymmetry (unit cell contains elements of different sizes), it can be observed in the spectrum as a resonant state with the finite quality factor depending on the level of asymmetry. We show that introducing loss and gain into the system allows us to control the system's response in both symmetric and asymmetric structures. In the symmetric case, the trapped mode can be excited for elements with almost equal loss and gain. In the asymmetric case, the detrimental impact of geometric dissimilarity can be compensated with the addition of the proper loss and gain to the elements of the structure, thus, restoring the high value of the quality factor. Our results demonstrate the capabilities of externally controlled gain media to vary the properties of optical resonances which is extremely important for applications in sensing, nonlinear optics, and laser physics.

\section{Geometry of an all-dielectric metamaterial}

\begin{figure}[t!]
\centering\includegraphics[width=\linewidth]{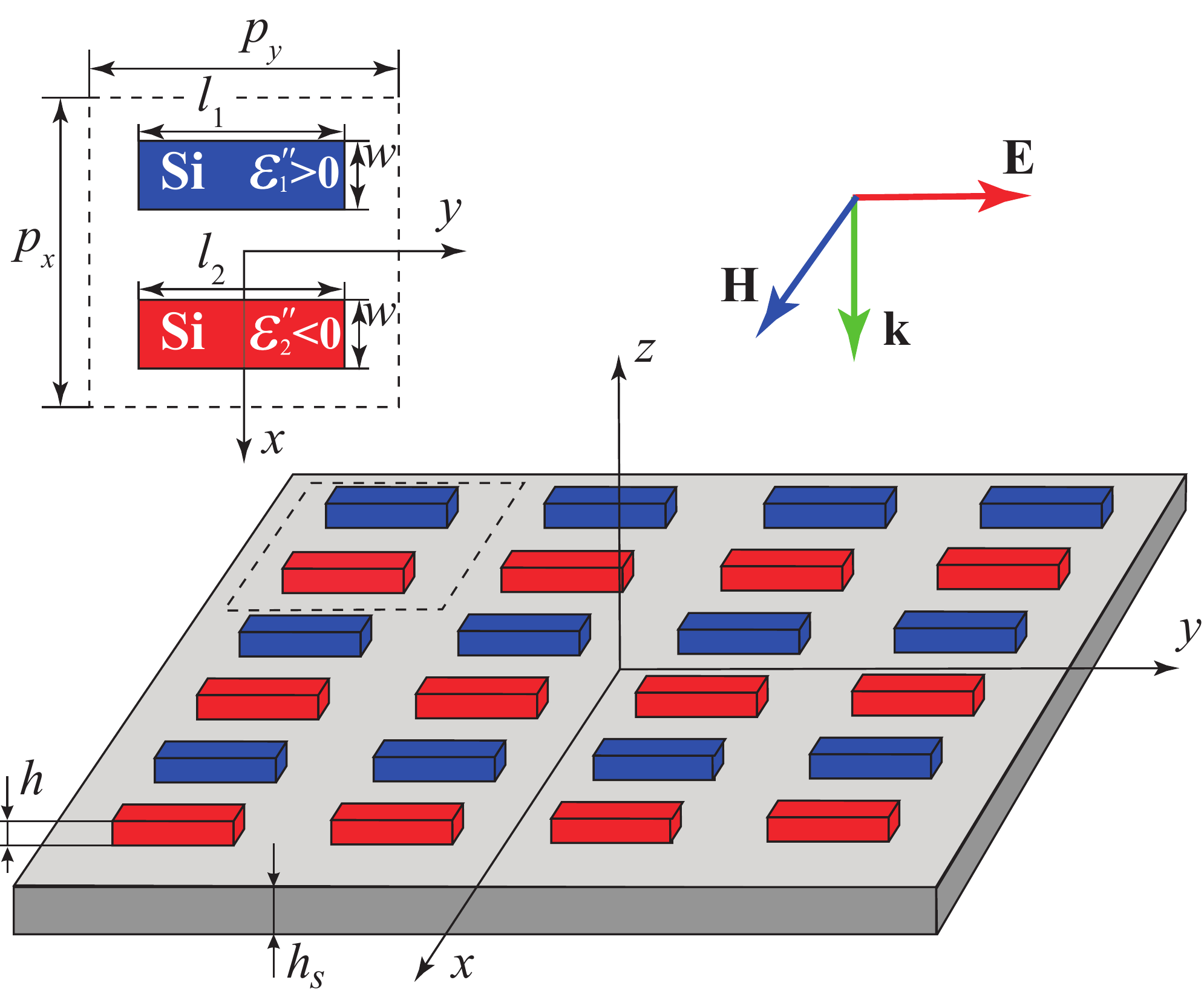}
\caption{\label{meta_surf} Fragment of an all-dielectric metamaterial and its elementary unit cell. Blue and red bars are made of material with loss (Si) $\varepsilon_1''>0$ and with gain (doped Si) $\varepsilon_2''<0$, respectively.}
\end{figure}

\begin{figure*}[t!]
\centering\includegraphics[width=\linewidth]{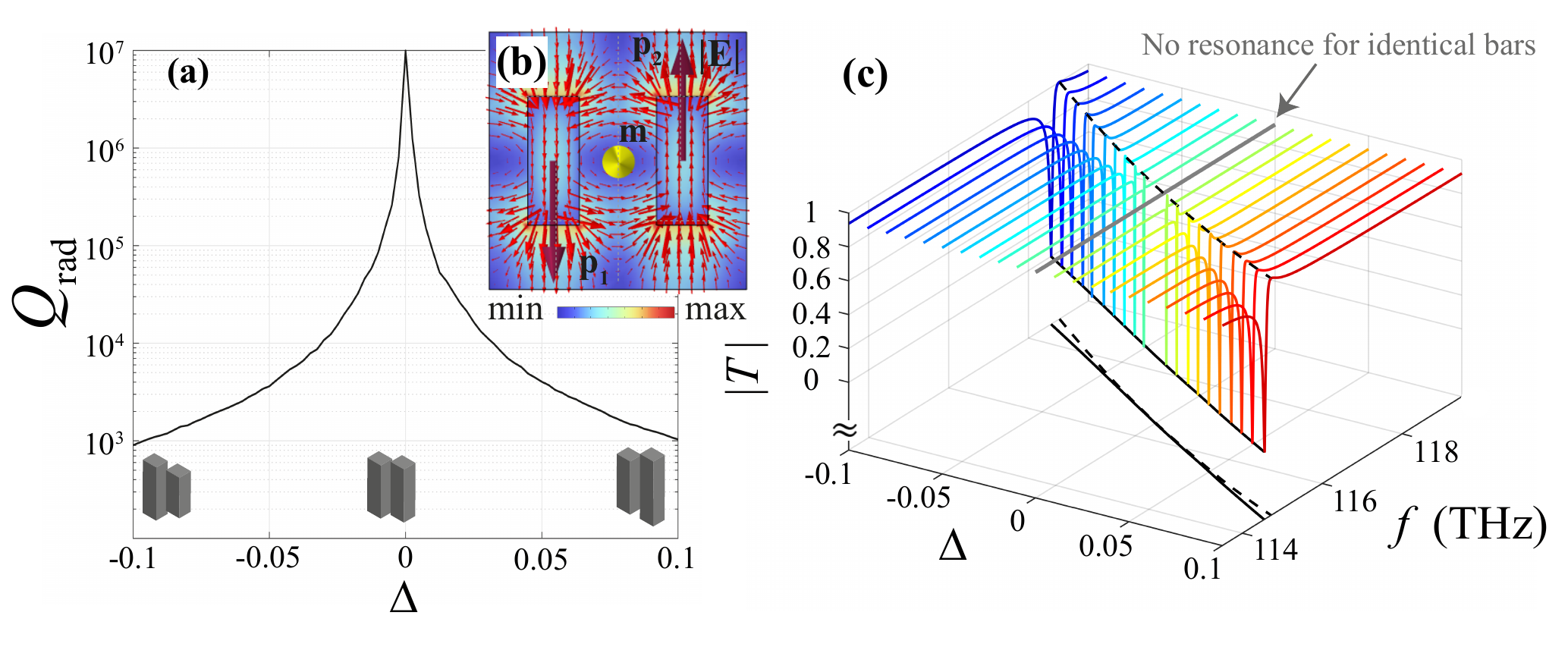}
\caption{\label{length_full} (a) Dependence of the quality factor ($Q_{\text{tot}}=Q_{\text{rad}}=\omega^\prime/2\omega^{\prime\prime}$, where $\omega$ is the complex eigenfrequency) of the eigenmode plotted on the logarithmic scale versus the asymmetry parameter $\Delta$, governing the difference between the lengths of the bars, $l_2=l_1(1+\Delta)$. (b) Distribution of the electric field magnitude corresponding to the trapped mode, where the red, brown, and yellow arrows represent the electric field strength flow and orientations of the electric dipole ($\textbf{p}_{1,2}$) and magnetic dipole ($\textbf{m}$) vectors, respectively. (c) Transmission coefficient magnitude $|T|$ versus the frequency $f=\omega/2\pi$ and asymmetry parameter $\Delta$. In the bottom plane, the positions of the peak and trough of the Fano-shape resonance are depicted.}
\end{figure*} 

In what follows, we consider an all-dielectric planar metamaterial consisting of a double-periodic grating shown in Fig. \ref{meta_surf}. The unit cell of this metamaterial contains a pair of optical dielectric nanoantennas made in the form of rectangular bars with permittivities $\varepsilon_1=\varepsilon_1'-i\varepsilon_1''$ and $\varepsilon_2=\varepsilon_2'-i\varepsilon_2''$, which are generally complex quantities taking a loss and gain into account (the field factor of the form $\exp(i\omega t)$ is assumed throughout the paper). The form of the unit cell is chosen to be square ($a = a_x = a_y$) and each unit cell is symmetric relative to the $x$-axis drawn through its center (see Fig. \ref{meta_surf}). The metamaterial is placed in an infinite homogeneous medium with real permittivity $\varepsilon_3 \in\Re$. In this study, air is used with $\varepsilon_3=1$, so that Re$(\varepsilon_{1,2}) > \varepsilon_3$. For our system, the values $\varepsilon''>0$ and $\varepsilon''<0$ correspond to media with loss and gain, respectively. It is assumed that permeability $\mu=\mu_0$ is the same for the overall system. In our configuration, the bars have equal thickness $h=h_1=h_2$ and the grating is placed on a thin dielectric substrate with permittivity $\varepsilon_s$ and thickness $h_s$. We assume that the structure is illuminated by a normally incident plane wave ($\textbf{k} = \{0,0,-k_z\}$) with the electric field polarized along the bars ($\textbf{E} = \{0,E_y,0\}$, $y$-polarization).

The length of the bars is limited by the size of the unit cell, which must be less than the wavelength of the incident field, and the thickness of the bars must be chosen so as to avoid the appearance of interference resonances. A typical high-index dielectric material for artificial nanostructures in the infrared region is silicon \cite{weber2018handbook, jahani2016all}. It is also necessary to provide high contrast between permittivities of the bars and substrate; therefore, at the first stage of our study, it is assumed that the substrate is made of a material with air-like permittivity, namely $\varepsilon_{s}=1$. This allows us to neglect the effects of substrate for a while and focus on the properties of the metamaterial itself. Then, in the final paragraphs of our study, the influence of the substrate used in practice is taken into account.

A perfect trapped mode has the infinite quality ($Q$) factor and cannot be observed in the metamaterial spectra under normal irradiation conditions. To excite this mode and observe it as a finite-$Q$ resonance, the system should be additionally modified making its unit cell asymmetric. Usually, asymmetry is introduced into the system, e.g., through a change in the geometry of the bars or their rotation \cite{khardikov2012giant, zhang2013near, tuz2018high, yu2021dielectric, dong2022nanoscale}. In this work, as a basic design, we excite the trapped mode by changing the size of one of the bars. The shape of the cell and bars is chosen to minimize the interaction between the modes. In our case, the trapped mode appears at a lower frequency than the other modes in the system. 

We start with the symmetric non-dissipative (lossless) system with the following parameters: the metamaterial period $a=2~\mu\text{m}$; the bar widths $w=w_1=w_2=0.4~\mu\text{m}$, lengths $l_1=l_2=1~\mu\text{m}$ and heights $h=0.3~\mu\text{m}$; bars permittivities $\varepsilon_1=\varepsilon_2=12$ are close to that of silicon; the distance between the center of bars is equal to half the cell size $d=1~\mu\text{m}$. To introduce asymmetry, we fix the size of both the unit cell and the first bar $l_1$ and vary the length of the second bar $l_2$. Thus, initially, we consider a symmetric unit cell configuration in which both bars are of the same size, and then decrease or increase the length $l_2=l_1(1+\Delta)$ of a particular bar, where $\Delta \in [-0.1, 0.1]$. For all our subsequent calculations we use the COMSOL Multiphysics electromagnetic solver.

In general, the total quality factor ($Q_{\text{tot}}$) of the system can be expressed as the sum of terms related to radiative and dissipative (material) losses ($Q^{-1}_{\text{tot}} = Q^{-1}_{\text{rad}} + Q^{-1}_{\text{dis}}$). Since in this section, we study the metamaterial without dissipative losses ($\varepsilon_1^{\prime\prime}=\varepsilon_2^{\prime\prime}=0$), the quality factor of the overall system depends only on the degree of radiative losses ($Q_{\text{tot}}=Q_{\text{rad}}$) arising in the asymmetric unit cells. 

The results of our calculations for the chosen parameters of the metamaterial are shown in Fig. \ref{length_full}. From the eigenmode analysis one can conclude that the resonant state of our interest has an infinite quality factor for the symmetric case whereas it becomes finite as soon as the asymmetry is introduced into the geometry of the unit cell of the structure ($|\Delta| \neq 0$) [Fig.~\ref{length_full}(a)]. 

The trapped mode under study is a resonance that arises through the coupling of closely spaced dielectric bars \cite{khardikov2012giant}. It appears from the antiparallel dipolar eigenstate characterized by a pair of in-plane electric dipole vectors $\textbf{p}_1$ and $\textbf{p}_2$ as illustrated schematically in Fig.~\ref{length_full}(b). For such an eigenstate, a magnetic dipole moment $\textbf{m}$ appears to be oriented out-of-plane. As long as the bars are identically placed and parallel, the electric dipoles are strictly antiparallel ($\textbf{p}_1 = -\textbf{p}_2$), and the resonance in the spectrum is not observed. Breaking the unit cell in-plane symmetry allows the mode coupling to the incident wave, resulting in the resonance arising as shown in Fig.~\ref{length_full}(c). 

\begin{figure*}[t!]
\centering\includegraphics[width=\linewidth]{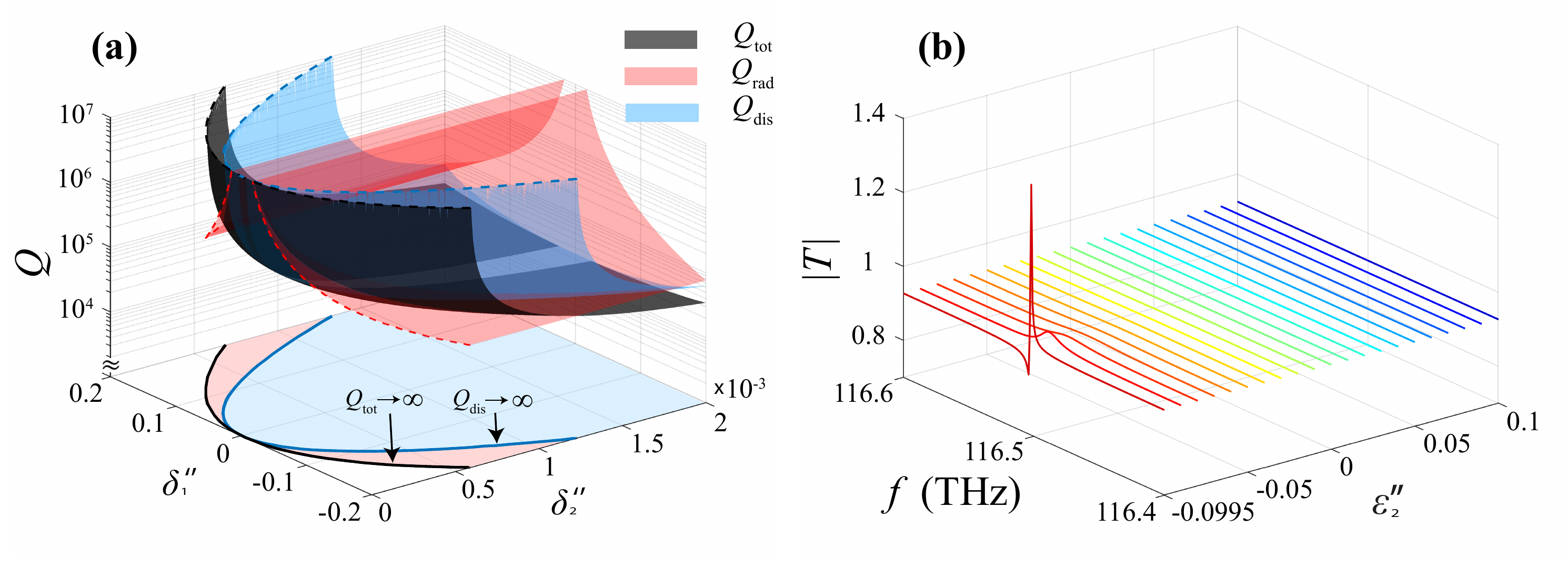}
\caption{\label{BIC_gain_loss} (a) Total ($Q_{\text{tot}}$), dissipative ($Q_{\text{dis}}$) and radiative ($Q_{\text{rad}}$) quality factors of the eigenmode plotted on the logarithmic scale versus the parameters $\delta_1''=\sqrt{2}(\varepsilon_1''-\varepsilon_2'')/2$ and $\delta_2''=\sqrt{2}(\varepsilon_1''+\varepsilon_2'')/2$. In the bottom plane, the position of the infinite values of $Q_{\text{tot}}$ (black line) and $Q_{\text{dis}}$ (blue line) are shown. The blue and red regions on the bottom plane correspond to conditions $Q_{\text{dis}}>0$ and $Q_{\text{dis}}<0$. (b) The transmission coefficient magnitude $|T|$ versus frequency $f=\omega/2\pi$ and the imaginary part of permittivity $\varepsilon_{2}''$.}
\end{figure*} 

The corresponding resonance has a Fano profile with the sharp peak and trough corresponding to transmission and reflection maxima, respectively, as is typical for the trapped modes excitation \cite{fedotov2007sharp, tuz2018high}. For the system without dissipative losses, the peak and trough tend to be $0$ or $1$. The smaller the asymmetry in the structure geometry, the lower the radiative losses and the higher the quality factor of the resonance. For the Fano resonance, the quality factor can be also associated with the distance between the frequency positions of the peak and trough: the large distance means the decrease in the quality factor of the resonance. An illustration of this feature can be seen in Fig. \ref{length_full}(c), where the black dashed and solid lines are the projections of maximum and minimum on the plane $(f,\Delta)$. Moreover, it is known \cite{Koshelev2018PRL} that the radiative part of the quality factor for such resonance is connected to the asymmetry parameter as $Q_{\text{rad}} \sim \Delta^{-2}$. One can see that this is the case for the considered metamaterial as well [Fig. \ref{length_full}(a)].

Thus, we can strongly manipulate the optical response of the system by changing its geometry close to the resonant state. In our case, the asymmetry of the system leads to the appearance of the resonance, which cannot be excited in the symmetric case. Further, we give the non-Hermitian generalization of this analysis.

\section{All-dielectric metamaterial with loss and gain}

Now we consider a metamaterial whose unit cell consists of a pair of dielectric rectangular bars with loss and gain. Here permittivities are complex quantities, where one of the bars (see Fig. \ref{meta_surf}) is made of a medium with loss (the blue bar with $\varepsilon''_{1}>0$) whereas the second one contains a material with gain (the red bar with $\varepsilon''_{2}<0$; we suppose that as a gain material, erbium-doped silicon \cite{Xie_JApplPhys_1991, Thao_JApplPhys_2000} can be used). The real part of permittivity for both bars is the same as previously. The geometry of the system is kept the same as in the previous section so that the structure supports the trapped mode in the same frequency range. Without loss of generality, one can shift the operating frequency by changing the unit cell size: when the cell size decreases, the operating frequency increases.

In the usual $\cal{PT}$-symmetric systems, the loss is compensated by the gain due to the interaction between the modes associated with each resonator \cite{ozdemir2019parity, hlushchenko2021multimode}. In the system under consideration, we are dealing with a trapped mode which is the mode of the entire unit cell of the structure, see the corresponding distribution of electromagnetic fields in Fig. \ref{length_full}(b). Since it is spectrally separated from the other modes, we are not able to obtain $\cal{PT}$ symmetry with the trapping mode alone. However, as we show here, we can achieve loss compensation in the metamaterial with loss and gain. To do this, we vary the imaginary parts of the permittivity of each bar. For better visualization of the results, we rotate the $(\varepsilon_1,\varepsilon_2)$ plane by an angle $-\pi/4$ and introduce a new coordinate system: $\delta_1''=\sqrt{2}(\varepsilon_1''-\varepsilon_2'')/2$, $\delta_2''=\sqrt{2}(\varepsilon_1''+\varepsilon_2'')/2$. Since $\varepsilon_1''$ and $\varepsilon_2''$ have different signs in the loss-gain system, the first coordinate $\delta_1''$ shows the total level of non-Hermiticity. In contrast, the second coordinate $\delta_2''$ is the difference between loss and gain magnitudes.

\begin{figure*}[t!]
\centering\includegraphics[width=\linewidth]{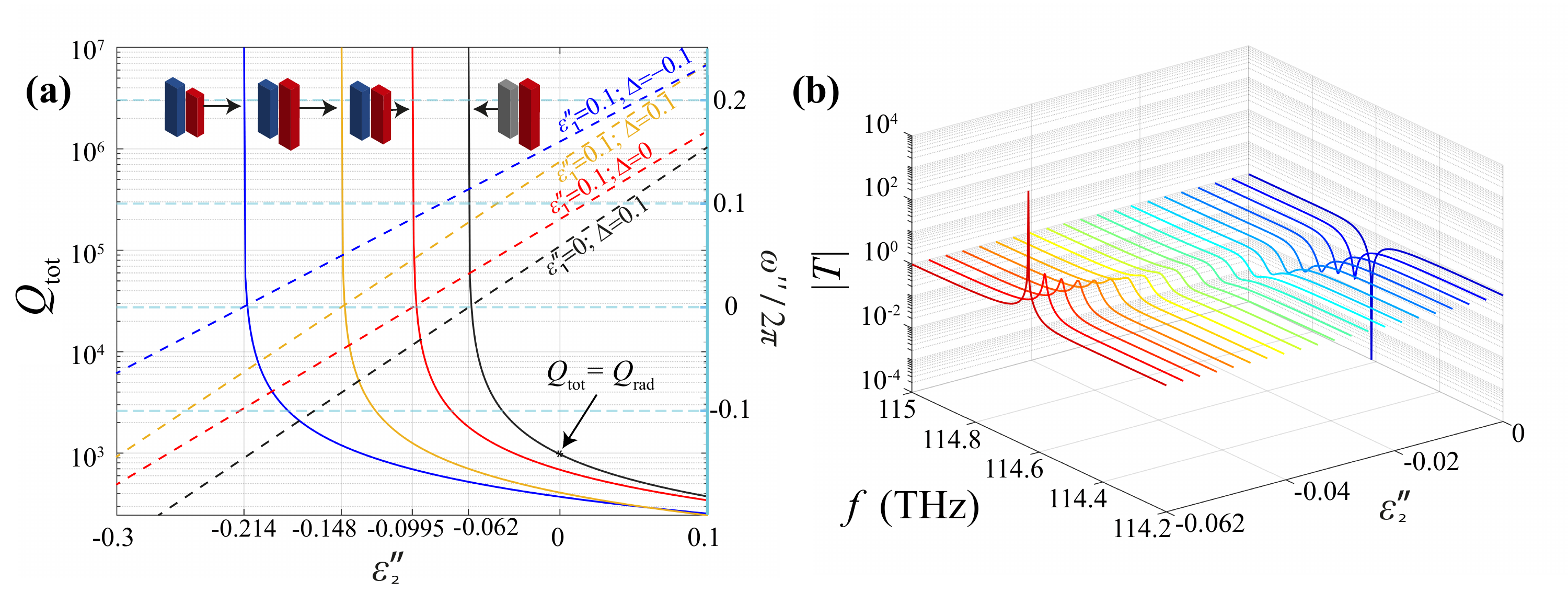}
\caption{\label{length_gain_loss} (a) Total quality factor (solid curves) and imaginary part of eigenfrequency $\omega''/2\pi$ (dashed lines)  for the asymmetric structure ($\Delta=\{-0.1,0,0.1\}$) versus the imaginary part of permittivity $\varepsilon_{2}''$. The asterisk corresponds to the quality factor of an asymmetric metamaterial ($\Delta=0.1$) with only radiative losses ($\varepsilon_{1}'' = \varepsilon_{2}'' = 0$). (b) The transmission coefficient magnitude $|T|$ versus frequency $f=\omega/2\pi$ and imaginary part of permittivity $\varepsilon_{2}''$ (gain). Here both the quality factor and transmission coefficient are plotted on the logarithmic scale.}
\end{figure*} 

As soon as the loss and gain are introduced to the system, both radiative $Q_\textrm{rad}$ and dissipative $Q_\textrm{dis}$ contributions to the quality factor should be accounted for. The contribution of dissipative losses can be calculated using the formula \cite{krupka2005extremely,shcherbinin2021IMTW}
\begin{equation}
\label{eq:q_dis}
Q^{-1}_{\text{dis}}=\sum\limits_{i=1}^2 \xi_{i}\tan{\delta}_i,
\end{equation}
where $\xi_{i}$ is the electric energy filling factor of the $i$-th bar and $\tan{\delta}_i=\varepsilon_i''/\varepsilon_i'$. The value of $\xi_{i}$ can be found as 

\begin{equation}
\xi_{i}=\frac{\iiint\limits_{V_{\textrm{bar},i}}\varepsilon_i'|\textbf{E}|^2d^3v}{\iiint\limits_{V_{\textrm{total}}}\varepsilon'(v)|\textbf{E}|^2d^3v}.
\label{eq:xi_dis}
\end{equation}


We start with the trapped mode characterization in the metamaterial having the bars with identical geometric parameters. The total, radiative, dissipative quality factors and manifestation of the mode in the transmitted spectra of the structure are presented in Fig. \ref{BIC_gain_loss}. 

The peak of $Q_{\text{tot}}$ [see the black surface in Fig. \ref{BIC_gain_loss}(a)] changes non-linearly with the parameters $\delta''_1$ and $\delta''_2$ (the projection of this relationship on the parameter plane at the bottom of Fig. \ref{BIC_gain_loss}(a) is shown by the black line). One can see that the trapped mode has the maximal $Q_{\text{tot}}$ for various loss and gain balance in the bars. When passing through the limit $Q_{\text{tot}}\rightarrow\infty$, the value of the total quality factor changes sign from plus to minus. After this threshold, the definition of quality factor is no longer applicable, since the gain becomes greater than the losses in the system. The blue surface in Fig. \ref{BIC_gain_loss}(a) corresponds to the dissipative losses $Q_{\text{dis}}$, which are calculated by Eq. (\ref{eq:q_dis}). As the gain in one of the bars increases, the compensation of dissipative losses appears, where the full compensations is at $Q_{\text{dis}}\rightarrow\infty$ [see the blue line on the bottom plane in Fig. \ref{BIC_gain_loss}(a)]. Compensation of dissipative losses entails an increase in radiation losses, which is associated with a decrease in $Q_{\text{rad}}$ [see the red surface in Fig. \ref{BIC_gain_loss}(a)]. The resulting radiation losses can be compensated by a further increase in the gain in one of the bars up to the values $Q_{\text{tot}}\rightarrow\infty$ [see the red area on the bottom plane in Fig. \ref{BIC_gain_loss}(a)]. 

Therefore, we can control the quality factor of the trapped mode simply by changing the level of loss and gain: for the fixed value of loss in one of the bars, we can restore the quality factor of the mode compensating both radiative and dissipative losses by adjusting the gain in the other bar. From a practical point of view, controlling an optical system by tuning its active (gain) elements is preferable for many applications than changing the geometry. Due to the huge technological progress in introducing gain elements into optical systems \cite{ozdemir2019parity,peng2014loss,sugioka2017progress,weber2018handbook}, such possibility is now more accessible and easier to control.

We plot in Fig. \ref{BIC_gain_loss}(b) the transmission coefficient magnitude as a function of frequency for the fixed loss in the first bar ($\varepsilon_1''=0.1$) and variable gain in the second bar ($\varepsilon_2'' \in [-0.2, 0.1]$). One can see that the trapped mode excitation can be controlled by changing the relationship between loss and gain and that the resonance occurs only close to $\varepsilon_2''=-0.0995$, i.e., for the loss and gain close in their magnitude. Due to the fact that our system contains active constituents, the maximum transmission coefficient can be greater than unity.

\begin{figure*}[t!]
\centering\includegraphics[width=\linewidth]{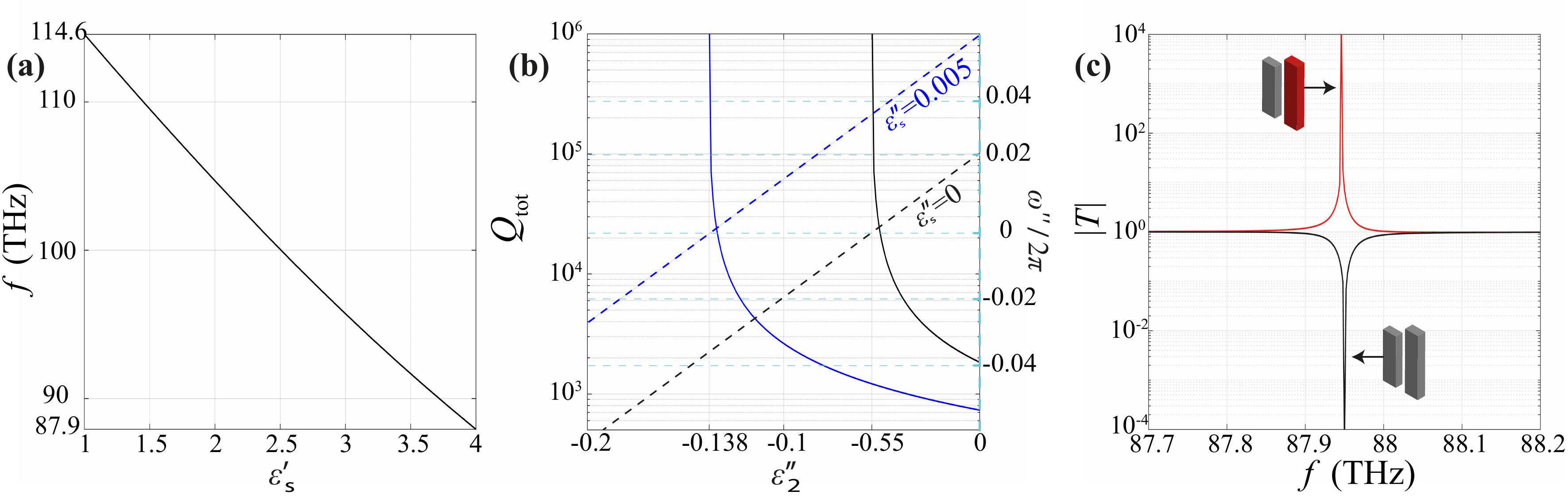}
\caption{\label{fig_5} (a) Dispersion curve versus the real part of substrate permittivity $\varepsilon_{s}$. (b) Total quality factor (solid curves) and imaginary part of eigenfrequency $\omega''/2\pi$ (dashed lines) for the metamaterial with substrate ($\varepsilon_{s}'' = \{0,0.005\}$) versus the imaginary part of the bar permittivity $\varepsilon_{2}''$. (c) The transmission coefficient magnitude $|T|$ versus frequency $f=\omega/2\pi$ ($\varepsilon_{s}=4$). Here both the quality factor and transmission coefficient are plotted on the logarithmic scale.}
\end{figure*} 

The most important point is that the trapped mode can be excited in the metamaterial by introducing gain without the need for breaking geometric symmetry, and the quality factor of this resonance is completely controlled by the active elements of the system. However, the experimental realization of this resonance requires very fine tuning of loss and gain in the bars that complicate its practical realization. Therefore, we further consider an asymmetric structure with dissimilar bars and show that full loss compensation can be achieved by controlling the gain. To restore the quality factor of the perfect trapped mode in an asymmetric structure, both radiative and dissipative losses need to be compensated. First, in order to estimate the gain required to compensate for radiative losses, we consider the bars of the dissimilar length ($l_1=1.155~\mu\text{m}$, $\Delta=0.1$). There is no dissipation in one of the bars ($\varepsilon^{\prime\prime}_1=0$), whereas we introduce either loss or gain into another bar ($\varepsilon_2'' \in[-0.3,0.1]$). The total quality factor and imaginary part of the eigenfrequency for this case are plotted in Fig. \ref{length_gain_loss}(a) by solid and dashed black lines, respectively. The value of the quality factor of the corresponding non-dissipative structure ($\varepsilon^{\prime\prime}_1=\varepsilon^{\prime\prime}_2=0$, $Q_{\text{tot}}=Q_{\text{rad}}$) accounting for radiative losses only is marked in this figure by an asterisk. Next, we introduce dissipation (loss) to the first bar  ($\varepsilon_1''=0.1$) and calculate the same parameters for three different lengths of the second bar. The results of our calculations are shown in Fig. \ref{length_gain_loss}(a) with color lines. 

As we have discussed above, the trapped mode can be easily excited in the asymmetric system, but the quality factor rapidly decreases with the growing asymmetry. This decrease can be compensated in the presence of loss and gain, which can be tuned so that the quality factor grows again. It is revealed that for the chosen geometric parameters of asymmetry used in our calculations, we can fully compensate both radiative and dissipative losses. In particular, the compensation of radiative losses for the trapped mode is achieved at the gain value $\varepsilon_2''=-0.062$ whereas the full loss compensation for the rest considered cases is at $\varepsilon_2'' = \{-0.214, -0.0995, -0.148\}$, respectively. The values of loss compensation are formed by the degeneracy of the imaginary part of eigenfrequency $\omega''$. Changing the sign of $\omega''$  from plus to minus leads to the system transition from attenuation to amplification due to the form of the field factor $\exp(i\omega t)$. This shows that non-Hermiticity is an additional degree of freedom allowing control of high-quality resonances in asymmetric metastructures.
 
Manifestation of the trapped mode resonance in the transmitted spectra of the asymmetric metamaterial with a gain is presented in Fig. \ref{length_gain_loss}(b). This characteristic is calculated for the metamaterial without dissipation ($\varepsilon^{\prime\prime}_1=0$, $\Delta=0.1$). One can see that as soon as the gain is introduced ($\varepsilon^{\prime\prime}_2\in[-0.1,0]$), the Fano resonance undergoes a change in shape, realizing in the trough smoothing and the peak increasing. The resonance reaches its maximum value at some optimal gain ($\varepsilon_2''=-0.062$) which corresponds to the point on the $\varepsilon_2''$ scale where the total quality factor tends to infinity. Thus, by controlling the asymmetry of the system and the value of introduced gain, one can control the distance between extremes of the Fano resonance.

Finally, we reveal how the substrate presence affects the trapped mode manifestation. In this study we fix the substrate thickness equal to two heights of the bars ($h_s=2h=0.6$ $\mu$m) and vary both real ($\varepsilon^{\prime}_s \in [1.0,4.0]$) and imaginary ($\varepsilon^{\prime\prime}_s = \{0,0.005\}$) parts of its permittivity. The results of our calculations are summarized in Fig. \ref{fig_5}. 

In particular, the resonant frequency of the trapped mode decreases as $\varepsilon^\prime_s$ increases [Fig. \ref{fig_5}(a)]. An optimal value of $\varepsilon^{\prime\prime}_2$ for loss compensation acquires some shift [Fig. \ref{fig_5}(b)], demonstrating that the losses inherent in the substrate can be also compensated. The characteristic of the transmission coefficient magnitude generally remains unchanged [Fig. \ref{fig_5}(c)], although the peak and trough of the Fano resonance may reverse their positions on the frequency scale. In total, the substrate presence does not perturb the in-plane symmetry of the structure and thus does not affect the characteristics of this particular trapped mode. To take into account the influence of the substrate, just a slight adjustment of the metamaterial parameters is required.

\section{Coupled-oscillator model}

In this section, we show that the numerical results described above can be directly interpreted with the simple model of two coupled non-Hermitian oscillators. Each oscillator substitutes the bar of the metamaterial discussed in the previous sections. The model is based on a pair of equations for two electric dipole moments $P_1$ and $P_2$:
\begin{eqnarray}
\ddot{P}_1 + \gamma_1 \dot{P}_1 + \omega^2_1 P_1 + c P_2 = A_1 E, \\
\ddot{P}_2 + \gamma_2 \dot{P}_2 + \omega^2_2 P_2 + c P_1 = A_2 E,
\end{eqnarray}
where $\omega_{1,2}$ are the resonant frequencies of both oscillators, $\gamma_{1,2}$ are their damping rates taking into account loss or gain, $A_{1,2}$ are the parameters governing interaction with the external electric field $E$, and $c$ is the coupling strength which is supposed to be the same for both oscillators due to reciprocity. Introducing slowly-varying amplitudes through $P_i=p_i e^{i \omega t}$ and $E=E_0 e^{i \omega t}$ and neglecting the second-order derivatives, we come to the first-order differential equations which can be conveniently written in the matrix form:
\begin{eqnarray}
-i \dot{\Pi} = \hat{H} \Pi + \Psi E_0, \label{matreq}
\end{eqnarray}
where $\Pi=(p_1, p_2)^T$, $\Psi=(\alpha_1, \alpha_2)^T$; $\hat{H}=\left( \begin{array}{cc} {\Omega_1} & {\kappa_1} \\ {\kappa_2} & {\Omega_2} \end{array} \right)$, $\Omega_j=(\omega^2_j - \omega^2 + i \omega \gamma_j)/(2 \omega - i \gamma_j)$, $\kappa_j=c/(2 \omega - i \gamma_j)$, $\alpha_j=-A_j/(2 \omega - i \gamma_j)$, $j=1,2$.

In absence of an external field ($E_0=0$), system (\ref{matreq}) takes the form of the Schr\"{o}dinger equation with the matrix $\hat{H}$ having the meaning of effective Hamiltonian. It is worth to find the eigenvalues and eigenvectors of this Hamiltonian, i.e., to solve the equation $\hat{H} \Phi=\lambda \Phi$ assuming $\Pi=\Phi e^{i \lambda t}$. The result is
\begin{eqnarray}
\lambda_\pm &=& \frac{1}{2} \left[ \Omega_1+\Omega_2 \pm \sqrt{4 \kappa_1 \kappa_2 + (\Omega_1-\Omega_2)^2} \right], \label{eigenvalues} \\
\Phi_\pm &=& \left( 1, \frac{\lambda_\pm - \Omega_1}{\kappa_1} \right)^T. \label{eigenvectors}
\end{eqnarray}
In general, the external field $E_0 \neq 0$ will excite both eigenmodes (\ref{eigenvectors}), so we can write the solution of Eq. (\ref{matreq}) as their linear combination, $\Pi(t)=\beta_+(t) \Phi_+ e^{i \lambda_+ t} + \beta_-(t) \Phi_- e^{i \lambda_- t}$. The corresponding differential equations for the coefficients $\beta_\pm (t)$ are as follows:

\begin{eqnarray}
\dot{\beta}_+ &=& i \frac{\alpha_1 (\Omega_1 - \lambda_-) + \alpha_2 \kappa_1}{\lambda_+ - \lambda_-} e^{-i \lambda_+ t} E_0, \label{eqbetaplus} \\
\dot{\beta}_- &=& -i \frac{\alpha_1 (\Omega_1 - \lambda_+) + \alpha_2 \kappa_1}{\lambda_+ - \lambda_-} e^{-i \lambda_- t} E_0. \label{eqbetaminus}
\end{eqnarray}
Supposing the stationary amplitude $E_0=\rm{const}$, we can solve these equations, so that
\begin{eqnarray}
\beta_+ (t) &=& \frac{\alpha_1 (\lambda_- - \Omega_1) - \alpha_2 \kappa_1}{\lambda_+ (\lambda_+ - \lambda_-)} (e^{-i \lambda_+ t} -1) E_0, \label{betaplus} \\
\beta_- (t) &=& \frac{\alpha_1 (\Omega_1 - \lambda_+) + \alpha_2 \kappa_1}{\lambda_- (\lambda_+ - \lambda_-)} (e^{-i \lambda_- t} -1) E_0. \label{betaminus}
\end{eqnarray}

\begin{figure}[t!]
\centering\includegraphics[width=\linewidth]{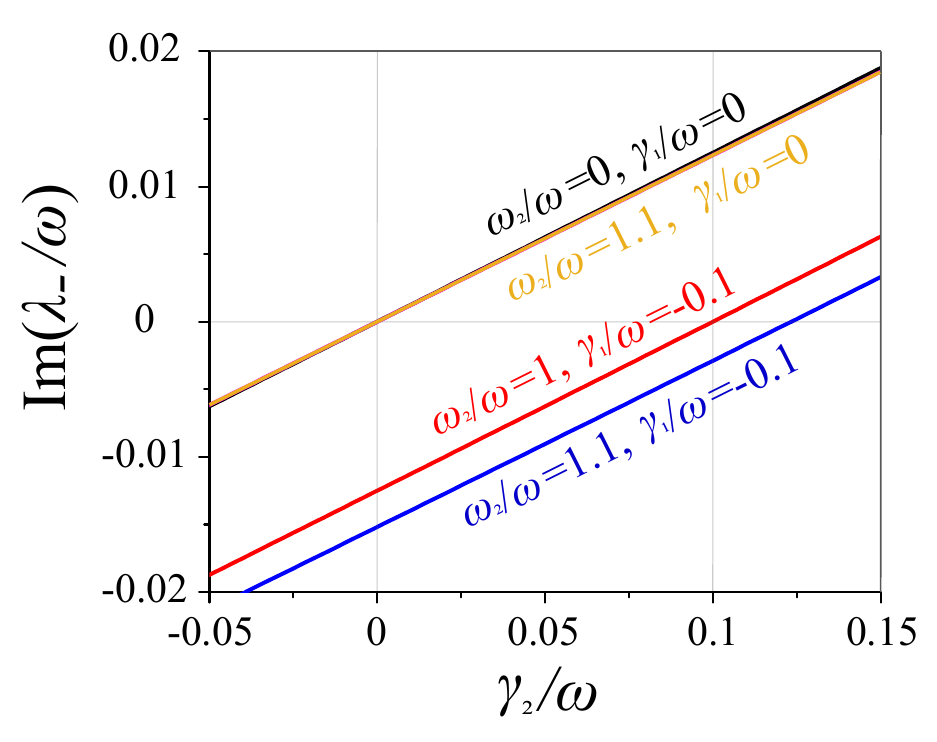}
\caption{\label{imag_lambda} Imaginary part of of the eigenvalue $\lambda_-$ versus the gain level $\gamma_2$. In calculations, we used $\omega_1/\omega=1$, $\alpha_1/\omega^2=\alpha_2/\omega^2=c/\omega^2=1$; the other parameters are shown on the figure.}
\end{figure}

These equations show changing contributions of the eigenmodes to the system's response. Returning to the dipoles, we obtain their dynamics:
\begin{eqnarray}
p_1(t)&=&\beta_+ (t) e^{i \lambda_+ t} + \beta_- (t) e^{i \lambda_- t}, \label{p1} \\
p_2(t)&=&\beta_+ (t) \frac{\lambda_+ - \Omega}{\kappa} e^{i \lambda_+ t} + \beta_- (t) \frac{\lambda_- - \Omega}{\kappa} e^{i \lambda_- t}. \label{p2}
\end{eqnarray}

To clearly demonstrate the meaning of the modes, at first, we consider the case of identical Hermitian oscillators with $\omega_1 = \omega_2 = \omega_0$, $A_1=A_2=A_0$, and $\gamma_1 = \gamma_2 = 0$. Then, all the auxiliary parameters are real: $\Omega=(\omega^2_0 - \omega^2)/(2 \omega)$, $\kappa=c/(2 \omega)$, $\alpha=-A_0/(2 \omega)$. For the eigenvalues and eigenvectors, we have $\lambda_\pm = \Omega \pm \kappa$ and $\Phi_\pm=(1,\pm 1)^T$. The first of these modes can be interpreted as the symmetric one (the dipoles oscillate in phase), whereas the second mode can be called the asymmetric one (the dipoles oscillate out of phase). Moreover, from Eqs. (\ref{betaplus}) and (\ref{betaminus}), we readily obtain $\beta_+ (t) = E_0 (1 - e^{-i \lambda_+ t}) \alpha_0/\lambda_+$ and $\beta_- (t) \equiv 0$. In other words, only the symmetric mode $\Phi_+$ contributes to the response of the system. The asymmetric mode $\Phi_-$ cannot be excited and can be considered as a trapped (dark) one.

If the oscillators are dissimilar or non-Hermitian, the asymmetric mode gets less dark and can show itself in the system's response. If we change the length of a bar in our metamaterial, then the resonant frequency of an oscillator and the other parameters in the model may change, so that the coefficient in Eq. (\ref{betaminus}) is not identically zero. The oscillations will then be governed by the exponential factor, which results in either attenuation or amplification depending on the imaginary part of the eigenvalue $\lambda_-$. We may suppose that the restoration of the trapped mode corresponds to the complete compensation of loss and gain when $\textrm{Im} \lambda_-=0$. Since the imaginary part of eigenfrequency governs the resonance width, this condition corresponds to the maximum quality factor. In Fig. \ref{imag_lambda}, we show the behavior of $\textrm{Im} \lambda_-$ as a function of $\gamma_2$ for several typical cases. The asymmetry is introduced into the model by changing the oscillator frequencies so that $\omega_2 \neq \omega_1$. One can see that when the first oscillator is Hermitian ($\gamma_1=0$), the second one should be Hermitian as well in order to avoid attenuation or amplification of oscillations. If we introduce loss into the system ($\gamma_1/\omega=-0.1$), one should add some gain to compensate for it and reach the condition $\textrm{Im} \lambda_-=0$. For the symmetric structure ($\omega_2 = \omega_1$), gain should be equal to loss for such compensation, $\gamma_2/\omega=-\gamma_1/\omega=0.1$. This is in accordance with the calculations in Fig. \ref{length_gain_loss} demonstrating almost equal imaginary parts of permittivity needed for the maximum quality factor in the symmetric case. For the asymmetric structure with $\omega_2 \neq \omega_1$, the condition $\textrm{Im} \lambda_-=0$ is reached at unequal gain and loss as shown in Fig. \ref{imag_lambda} for $\omega_2/\omega=1.1$. This situation corresponds to the dashed lines obtained numerically for the asymmetric case presented in Fig. \ref{length_gain_loss}.

Although the strict quantitative correspondence between the simple coupled-oscillator model and the numerical calculations cannot be carried out, we believe that these considerations are helpful for a qualitative understanding the results obtained in this paper.

\section{Conclusion}

We have elucidated the role of active (gain) elements in controlling the optical response of a planar all-dielectric metamaterial (metasurface). In particular, the trapped mode in the structure with the symmetric elementary cell containing two identical bars with loss and gain can be controlled to achieve the Fano resonance in the transmission spectrum. Simultaneously, the quality factor can be tuned to its maximum value by varying loss and gain. In the asymmetric system with unequal bars, the geometric asymmetry can be compensated by tuning loss and gain to restore the high value of the quality factor characteristic for the trapped mode. Our results show how the quality factor of the resonance can be controlled with both geometry and non-Hermiticity variations significantly expanding the experimental possibilities for implementation of the tunable optical systems. In contrast to the structural transformations used for tuning the properties of metamaterials, the gain is an external factor that can be dynamically controlled by changing pump intensity. This is an undoubtedly preferable method of response tuning from the practical perspective.

\acknowledgements{The authors are grateful for the hospitality and support from Jilin University, China. This work was partially supported (A.V.H.) by the  Grant of the National Academy of Sciences (NAS) of Ukraine to research laboratories/groups of young scientists of the NAS of Ukraine for conducting research in priority areas of science and technology (Grant no. 0122U002145).}

\bigskip

\bibliography{trapped_mode}

\begin{thebibliography}{57}%
\makeatletter
\providecommand \@ifxundefined [1]{%
 \@ifx{#1\undefined}
}%
\providecommand \@ifnum [1]{%
 \ifnum #1\expandafter \@firstoftwo
 \else \expandafter \@secondoftwo
 \fi
}%
\providecommand \@ifx [1]{%
 \ifx #1\expandafter \@firstoftwo
 \else \expandafter \@secondoftwo
 \fi
}%
\providecommand \natexlab [1]{#1}%
\providecommand \enquote  [1]{``#1''}%
\providecommand \bibnamefont  [1]{#1}%
\providecommand \bibfnamefont [1]{#1}%
\providecommand \citenamefont [1]{#1}%
\providecommand \href@noop [0]{\@secondoftwo}%
\providecommand \href [0]{\begingroup \@sanitize@url \@href}%
\providecommand \@href[1]{\@@startlink{#1}\@@href}%
\providecommand \@@href[1]{\endgroup#1\@@endlink}%
\providecommand \@sanitize@url [0]{\catcode `\\12\catcode `\$12\catcode
  `\&12\catcode `\#12\catcode `\^12\catcode `\_12\catcode `\%12\relax}%
\providecommand \@@startlink[1]{}%
\providecommand \@@endlink[0]{}%
\providecommand \url  [0]{\begingroup\@sanitize@url \@url }%
\providecommand \@url [1]{\endgroup\@href {#1}{\urlprefix }}%
\providecommand \urlprefix  [0]{URL }%
\providecommand \Eprint [0]{\href }%
\providecommand \doibase [0]{https://doi.org/}%
\providecommand \selectlanguage [0]{\@gobble}%
\providecommand \bibinfo  [0]{\@secondoftwo}%
\providecommand \bibfield  [0]{\@secondoftwo}%
\providecommand \translation [1]{[#1]}%
\providecommand \BibitemOpen [0]{}%
\providecommand \bibitemStop [0]{}%
\providecommand \bibitemNoStop [0]{.\EOS\space}%
\providecommand \EOS [0]{\spacefactor3000\relax}%
\providecommand \BibitemShut  [1]{\csname bibitem#1\endcsname}%
\let\auto@bib@innerbib\@empty
\bibitem [{\citenamefont {Krasnok}\ and\ \citenamefont
  {Al{\`u}}(2020)}]{krasnok2020active}%
  \BibitemOpen
  \bibfield  {author} {\bibinfo {author} {\bibfnamefont {A.}~\bibnamefont
  {Krasnok}}\ and\ \bibinfo {author} {\bibfnamefont {A.}~\bibnamefont
  {Al{\`u}}},\ }\bibfield  {title} {\bibinfo {title} {Active nanophotonics},\
  }\href {https://doi.org/10.1109/JPROC.2020.2985048} {\bibfield  {journal}
  {\bibinfo  {journal} {Proc. IEEE}\ }\textbf {\bibinfo {volume} {108}},\
  \bibinfo {pages} {628} (\bibinfo {year} {2020})}\BibitemShut {NoStop}%
\bibitem [{\citenamefont {R{\"u}ter}\ \emph {et~al.}(2010)\citenamefont
  {R{\"u}ter}, \citenamefont {Makris}, \citenamefont {El-Ganainy},
  \citenamefont {Christodoulides}, \citenamefont {Segev},\ and\ \citenamefont
  {Kip}}]{ruter2010observation}%
  \BibitemOpen
  \bibfield  {author} {\bibinfo {author} {\bibfnamefont {C.~E.}\ \bibnamefont
  {R{\"u}ter}}, \bibinfo {author} {\bibfnamefont {K.~G.}\ \bibnamefont
  {Makris}}, \bibinfo {author} {\bibfnamefont {R.}~\bibnamefont {El-Ganainy}},
  \bibinfo {author} {\bibfnamefont {D.~N.}\ \bibnamefont {Christodoulides}},
  \bibinfo {author} {\bibfnamefont {M.}~\bibnamefont {Segev}},\ and\ \bibinfo
  {author} {\bibfnamefont {D.}~\bibnamefont {Kip}},\ }\bibfield  {title}
  {\bibinfo {title} {Observation of parity--time symmetry in optics},\ }\href
  {https://doi.org/10.1038/nphys1515} {\bibfield  {journal} {\bibinfo
  {journal} {Nat. Phys.}\ }\textbf {\bibinfo {volume} {6}},\ \bibinfo {pages}
  {192} (\bibinfo {year} {2010})}\BibitemShut {NoStop}%
\bibitem [{\citenamefont {{\"O}zdemir}\ \emph {et~al.}(2019)\citenamefont
  {{\"O}zdemir}, \citenamefont {Rotter}, \citenamefont {Nori},\ and\
  \citenamefont {Yang}}]{ozdemir2019parity}%
  \BibitemOpen
  \bibfield  {author} {\bibinfo {author} {\bibfnamefont {{\c{S}}.~K.}\
  \bibnamefont {{\"O}zdemir}}, \bibinfo {author} {\bibfnamefont
  {S.}~\bibnamefont {Rotter}}, \bibinfo {author} {\bibfnamefont
  {F.}~\bibnamefont {Nori}},\ and\ \bibinfo {author} {\bibfnamefont
  {L.}~\bibnamefont {Yang}},\ }\bibfield  {title} {\bibinfo {title}
  {Parity--time symmetry and exceptional points in photonics},\ }\href
  {https://doi.org/10.1038/s41563-019-0304-9} {\bibfield  {journal} {\bibinfo
  {journal} {Nat. Mater.}\ }\textbf {\bibinfo {volume} {18}},\ \bibinfo {pages}
  {783} (\bibinfo {year} {2019})}\BibitemShut {NoStop}%
\bibitem [{\citenamefont {Kang}\ \emph {et~al.}(2016)\citenamefont {Kang},
  \citenamefont {Chen},\ and\ \citenamefont {Chong}}]{kang2016chiral}%
  \BibitemOpen
  \bibfield  {author} {\bibinfo {author} {\bibfnamefont {M.}~\bibnamefont
  {Kang}}, \bibinfo {author} {\bibfnamefont {J.}~\bibnamefont {Chen}},\ and\
  \bibinfo {author} {\bibfnamefont {Y.~D.}\ \bibnamefont {Chong}},\ }\bibfield
  {title} {\bibinfo {title} {Chiral exceptional points in metasurfaces},\
  }\href {https://doi.org/10.1103/PhysRevA.94.033834} {\bibfield  {journal}
  {\bibinfo  {journal} {Phys. Rev. A}\ }\textbf {\bibinfo {volume} {94}},\
  \bibinfo {pages} {033834} (\bibinfo {year} {2016})}\BibitemShut {NoStop}%
\bibitem [{\citenamefont {Alaee}\ \emph {et~al.}(2018)\citenamefont {Alaee},
  \citenamefont {Gurlek}, \citenamefont {Christensen},\ and\ \citenamefont
  {Kadic}}]{alaee2018optical}%
  \BibitemOpen
  \bibfield  {author} {\bibinfo {author} {\bibfnamefont {R.}~\bibnamefont
  {Alaee}}, \bibinfo {author} {\bibfnamefont {B.}~\bibnamefont {Gurlek}},
  \bibinfo {author} {\bibfnamefont {J.}~\bibnamefont {Christensen}},\ and\
  \bibinfo {author} {\bibfnamefont {M.}~\bibnamefont {Kadic}},\ }\bibfield
  {title} {\bibinfo {title} {Optical force rectifiers based on {PT}-symmetric
  metasurfaces},\ }\href {https://doi.org/10.1103/PhysRevB.97.195420}
  {\bibfield  {journal} {\bibinfo  {journal} {Phys. Rev. B}\ }\textbf {\bibinfo
  {volume} {97}},\ \bibinfo {pages} {195420} (\bibinfo {year}
  {2018})}\BibitemShut {NoStop}%
\bibitem [{\citenamefont {Dong}\ \emph {et~al.}(2020)\citenamefont {Dong},
  \citenamefont {Hu}, \citenamefont {Wang}, \citenamefont {Jia}, \citenamefont
  {Zhang}, \citenamefont {Cao}, \citenamefont {Wang}, \citenamefont {Chen},
  \citenamefont {Fan}, \citenamefont {Jiang} \emph {et~al.}}]{dong2020loss}%
  \BibitemOpen
  \bibfield  {author} {\bibinfo {author} {\bibfnamefont {S.}~\bibnamefont
  {Dong}}, \bibinfo {author} {\bibfnamefont {G.}~\bibnamefont {Hu}}, \bibinfo
  {author} {\bibfnamefont {Q.}~\bibnamefont {Wang}}, \bibinfo {author}
  {\bibfnamefont {Y.}~\bibnamefont {Jia}}, \bibinfo {author} {\bibfnamefont
  {Q.}~\bibnamefont {Zhang}}, \bibinfo {author} {\bibfnamefont
  {G.}~\bibnamefont {Cao}}, \bibinfo {author} {\bibfnamefont {J.}~\bibnamefont
  {Wang}}, \bibinfo {author} {\bibfnamefont {S.}~\bibnamefont {Chen}}, \bibinfo
  {author} {\bibfnamefont {D.}~\bibnamefont {Fan}}, \bibinfo {author}
  {\bibfnamefont {W.}~\bibnamefont {Jiang}}, \emph {et~al.},\ }\bibfield
  {title} {\bibinfo {title} {Loss-assisted metasurface at an exceptional
  point},\ }\href {https://doi.org/10.1021/acsphotonics.0c01440} {\bibfield
  {journal} {\bibinfo  {journal} {ACS Photonics}\ }\textbf {\bibinfo {volume}
  {7}},\ \bibinfo {pages} {3321} (\bibinfo {year} {2020})}\BibitemShut
  {NoStop}%
\bibitem [{\citenamefont {Song}\ \emph {et~al.}(2021)\citenamefont {Song},
  \citenamefont {Odeh}, \citenamefont {Z{\'u}{\~n}iga-P{\'e}rez}, \citenamefont
  {Kant{\'e}},\ and\ \citenamefont {Genevet}}]{song2021plasmonic}%
  \BibitemOpen
  \bibfield  {author} {\bibinfo {author} {\bibfnamefont {Q.}~\bibnamefont
  {Song}}, \bibinfo {author} {\bibfnamefont {M.}~\bibnamefont {Odeh}}, \bibinfo
  {author} {\bibfnamefont {J.}~\bibnamefont {Z{\'u}{\~n}iga-P{\'e}rez}},
  \bibinfo {author} {\bibfnamefont {B.}~\bibnamefont {Kant{\'e}}},\ and\
  \bibinfo {author} {\bibfnamefont {P.}~\bibnamefont {Genevet}},\ }\bibfield
  {title} {\bibinfo {title} {Plasmonic topological metasurface by encircling an
  exceptional point},\ }\href {https://doi.org/10.1126/science.abj3179}
  {\bibfield  {journal} {\bibinfo  {journal} {Science}\ }\textbf {\bibinfo
  {volume} {373}},\ \bibinfo {pages} {1133} (\bibinfo {year}
  {2021})}\BibitemShut {NoStop}%
\bibitem [{\citenamefont {Wang}\ \emph {et~al.}(2021)\citenamefont {Wang},
  \citenamefont {Shen}, \citenamefont {Yu}, \citenamefont {Zou}, \citenamefont
  {Ouyang},\ and\ \citenamefont {Deng}}]{wang2021active}%
  \BibitemOpen
  \bibfield  {author} {\bibinfo {author} {\bibfnamefont {J.}~\bibnamefont
  {Wang}}, \bibinfo {author} {\bibfnamefont {Y.}~\bibnamefont {Shen}}, \bibinfo
  {author} {\bibfnamefont {X.}~\bibnamefont {Yu}}, \bibinfo {author}
  {\bibfnamefont {L.}~\bibnamefont {Zou}}, \bibinfo {author} {\bibfnamefont
  {S.}~\bibnamefont {Ouyang}},\ and\ \bibinfo {author} {\bibfnamefont
  {X.}~\bibnamefont {Deng}},\ }\bibfield  {title} {\bibinfo {title} {Active
  control of parity-time symmetry phase transition in terahertz metasurface},\
  }\href {https://doi.org/https://doi.org/10.1016/j.physleta.2021.127304}
  {\bibfield  {journal} {\bibinfo  {journal} {Phys. Lett. A}\ }\textbf
  {\bibinfo {volume} {400}},\ \bibinfo {pages} {127304} (\bibinfo {year}
  {2021})}\BibitemShut {NoStop}%
\bibitem [{\citenamefont {Yu}\ \emph {et~al.}(2021)\citenamefont {Yu},
  \citenamefont {Ma}, \citenamefont {Ouyang}, \citenamefont {Ghosh},
  \citenamefont {Luo}, \citenamefont {Pattanayak}, \citenamefont {Kaur},
  \citenamefont {Qiu}, \citenamefont {Belov},\ and\ \citenamefont
  {Li}}]{yu2021dielectric}%
  \BibitemOpen
  \bibfield  {author} {\bibinfo {author} {\bibfnamefont {J.}~\bibnamefont
  {Yu}}, \bibinfo {author} {\bibfnamefont {B.}~\bibnamefont {Ma}}, \bibinfo
  {author} {\bibfnamefont {A.}~\bibnamefont {Ouyang}}, \bibinfo {author}
  {\bibfnamefont {P.}~\bibnamefont {Ghosh}}, \bibinfo {author} {\bibfnamefont
  {H.}~\bibnamefont {Luo}}, \bibinfo {author} {\bibfnamefont {A.}~\bibnamefont
  {Pattanayak}}, \bibinfo {author} {\bibfnamefont {S.}~\bibnamefont {Kaur}},
  \bibinfo {author} {\bibfnamefont {M.}~\bibnamefont {Qiu}}, \bibinfo {author}
  {\bibfnamefont {P.}~\bibnamefont {Belov}},\ and\ \bibinfo {author}
  {\bibfnamefont {Q.}~\bibnamefont {Li}},\ }\bibfield  {title} {\bibinfo
  {title} {Dielectric super-absorbing metasurfaces via {PT} symmetry
  breaking},\ }\href {https://doi.org/10.1364/OPTICA.430893} {\bibfield
  {journal} {\bibinfo  {journal} {Optica}\ }\textbf {\bibinfo {volume} {8}},\
  \bibinfo {pages} {1290} (\bibinfo {year} {2021})}\BibitemShut {NoStop}%
\bibitem [{\citenamefont {Dong}\ \emph {et~al.}(2022)\citenamefont {Dong},
  \citenamefont {Mahfoud}, \citenamefont {Paniagua-Dom{\'\i}nguez},
  \citenamefont {Wang}, \citenamefont {Fern{\'a}ndez-Dom{\'\i}nguez},
  \citenamefont {Gorelik}, \citenamefont {Ha}, \citenamefont {Tjiptoharsono},
  \citenamefont {Kuznetsov}, \citenamefont {Bosman} \emph
  {et~al.}}]{dong2022nanoscale}%
  \BibitemOpen
  \bibfield  {author} {\bibinfo {author} {\bibfnamefont {Z.}~\bibnamefont
  {Dong}}, \bibinfo {author} {\bibfnamefont {Z.}~\bibnamefont {Mahfoud}},
  \bibinfo {author} {\bibfnamefont {R.}~\bibnamefont
  {Paniagua-Dom{\'\i}nguez}}, \bibinfo {author} {\bibfnamefont
  {H.}~\bibnamefont {Wang}}, \bibinfo {author} {\bibfnamefont {A.~I.}\
  \bibnamefont {Fern{\'a}ndez-Dom{\'\i}nguez}}, \bibinfo {author}
  {\bibfnamefont {S.}~\bibnamefont {Gorelik}}, \bibinfo {author} {\bibfnamefont
  {S.~T.}\ \bibnamefont {Ha}}, \bibinfo {author} {\bibfnamefont
  {F.}~\bibnamefont {Tjiptoharsono}}, \bibinfo {author} {\bibfnamefont {A.~I.}\
  \bibnamefont {Kuznetsov}}, \bibinfo {author} {\bibfnamefont {M.}~\bibnamefont
  {Bosman}}, \emph {et~al.},\ }\bibfield  {title} {\bibinfo {title} {Nanoscale
  mapping of optically inaccessible bound-states-in-the-continuum},\ }\href
  {https://doi.org/10.1038/s41377-021-00707-2} {\bibfield  {journal} {\bibinfo
  {journal} {Light Sci. Appl.}\ }\textbf {\bibinfo {volume} {11}},\ \bibinfo
  {pages} {1} (\bibinfo {year} {2022})}\BibitemShut {NoStop}%
\bibitem [{\citenamefont {Hlushchenko}\ \emph
  {et~al.}(2021{\natexlab{a}})\citenamefont {Hlushchenko}, \citenamefont
  {Shcherbinin}, \citenamefont {Novitsky},\ and\ \citenamefont
  {Tuz}}]{hlushchenko2021multimode}%
  \BibitemOpen
  \bibfield  {author} {\bibinfo {author} {\bibfnamefont {A.~V.}\ \bibnamefont
  {Hlushchenko}}, \bibinfo {author} {\bibfnamefont {V.~I.}\ \bibnamefont
  {Shcherbinin}}, \bibinfo {author} {\bibfnamefont {D.~V.}\ \bibnamefont
  {Novitsky}},\ and\ \bibinfo {author} {\bibfnamefont {V.~R.}\ \bibnamefont
  {Tuz}},\ }\bibfield  {title} {\bibinfo {title} {Multimode parity-time
  symmetry and loss compensation in coupled waveguides with loss and gain},\
  }\href {https://doi.org/10.1103/PhysRevA.104.013507} {\bibfield  {journal}
  {\bibinfo  {journal} {Phys. Rev. A}\ }\textbf {\bibinfo {volume} {104}},\
  \bibinfo {pages} {013507} (\bibinfo {year} {2021}{\natexlab{a}})}\BibitemShut
  {NoStop}%
\bibitem [{\citenamefont {Hlushchenko}\ \emph
  {et~al.}(2021{\natexlab{b}})\citenamefont {Hlushchenko}, \citenamefont
  {Novitsky}, \citenamefont {Shcherbinin},\ and\ \citenamefont
  {Tuz}}]{hlushchenko2021multimode2}%
  \BibitemOpen
  \bibfield  {author} {\bibinfo {author} {\bibfnamefont {A.~V.}\ \bibnamefont
  {Hlushchenko}}, \bibinfo {author} {\bibfnamefont {D.~V.}\ \bibnamefont
  {Novitsky}}, \bibinfo {author} {\bibfnamefont {V.~I.}\ \bibnamefont
  {Shcherbinin}},\ and\ \bibinfo {author} {\bibfnamefont {V.~R.}\ \bibnamefont
  {Tuz}},\ }\bibfield  {title} {\bibinfo {title} {Multimode-symmetry thresholds
  and third-order exceptional points in coupled dielectric waveguides with loss
  and gain},\ }\href {https://doi.org/10.1088/2040-8986/ac31d4} {\bibfield
  {journal} {\bibinfo  {journal} {J. Opt.}\ }\textbf {\bibinfo {volume} {23}},\
  \bibinfo {pages} {125002} (\bibinfo {year} {2021}{\natexlab{b}})}\BibitemShut
  {NoStop}%
\bibitem [{\citenamefont {Zheludev}\ and\ \citenamefont
  {Kivshar}(2012)}]{zheludev_NatureMat_2012}%
  \BibitemOpen
  \bibfield  {author} {\bibinfo {author} {\bibfnamefont {N.~I.}\ \bibnamefont
  {Zheludev}}\ and\ \bibinfo {author} {\bibfnamefont {Y.~S.}\ \bibnamefont
  {Kivshar}},\ }\bibfield  {title} {\bibinfo {title} {From metamaterials to
  metadevices},\ }\href {https://doi.org/10.1038/nmat3431} {\bibfield
  {journal} {\bibinfo  {journal} {Nat. Mater.}\ }\textbf {\bibinfo {volume}
  {11}},\ \bibinfo {pages} {917} (\bibinfo {year} {2012})}\BibitemShut
  {NoStop}%
\bibitem [{\citenamefont {Koenderink}\ \emph {et~al.}(2015)\citenamefont
  {Koenderink}, \citenamefont {Al{\`u}},\ and\ \citenamefont
  {Polman}}]{Koenderink_Scince_2015}%
  \BibitemOpen
  \bibfield  {author} {\bibinfo {author} {\bibfnamefont {A.~F.}\ \bibnamefont
  {Koenderink}}, \bibinfo {author} {\bibfnamefont {A.}~\bibnamefont
  {Al{\`u}}},\ and\ \bibinfo {author} {\bibfnamefont {A.}~\bibnamefont
  {Polman}},\ }\bibfield  {title} {\bibinfo {title} {Nanophotonics: {Shrinking}
  light-based technology},\ }\href {https://doi.org/10.1126/science.1261243}
  {\bibfield  {journal} {\bibinfo  {journal} {Science}\ }\textbf {\bibinfo
  {volume} {348}},\ \bibinfo {pages} {516} (\bibinfo {year}
  {2015})}\BibitemShut {NoStop}%
\bibitem [{\citenamefont {Neshev}\ and\ \citenamefont
  {Aharonovich}(2018)}]{Neshev_NatLight_2018}%
  \BibitemOpen
  \bibfield  {author} {\bibinfo {author} {\bibfnamefont {D.}~\bibnamefont
  {Neshev}}\ and\ \bibinfo {author} {\bibfnamefont {I.}~\bibnamefont
  {Aharonovich}},\ }\bibfield  {title} {\bibinfo {title} {Optical metasurfaces:
  new generation building blocks for multi-functional optics},\ }\href
  {https://doi.org/10.1038/s41377-018-0058-1} {\bibfield  {journal} {\bibinfo
  {journal} {Light Sci. Appl.}\ }\textbf {\bibinfo {volume} {7}},\ \bibinfo
  {pages} {1} (\bibinfo {year} {2018})}\BibitemShut {NoStop}%
\bibitem [{\citenamefont {Jahani}\ and\ \citenamefont
  {Jacob}(2016)}]{jahani2016all}%
  \BibitemOpen
  \bibfield  {author} {\bibinfo {author} {\bibfnamefont {S.}~\bibnamefont
  {Jahani}}\ and\ \bibinfo {author} {\bibfnamefont {Z.}~\bibnamefont {Jacob}},\
  }\bibfield  {title} {\bibinfo {title} {All-dielectric metamaterials},\ }\href
  {https://doi.org/10.1038/nnano.2015.304} {\bibfield  {journal} {\bibinfo
  {journal} {Nat. Nanotechnol.}\ }\textbf {\bibinfo {volume} {11}},\ \bibinfo
  {pages} {23} (\bibinfo {year} {2016})}\BibitemShut {NoStop}%
\bibitem [{\citenamefont {Gu}\ \emph {et~al.}(2012)\citenamefont {Gu},
  \citenamefont {Singh}, \citenamefont {Liu}, \citenamefont {Zhang},
  \citenamefont {Ma}, \citenamefont {Zhang}, \citenamefont {Maier},
  \citenamefont {Tian}, \citenamefont {Azad}, \citenamefont {Chen} \emph
  {et~al.}}]{Gu_NatureCom_2012}%
  \BibitemOpen
  \bibfield  {author} {\bibinfo {author} {\bibfnamefont {J.}~\bibnamefont
  {Gu}}, \bibinfo {author} {\bibfnamefont {R.}~\bibnamefont {Singh}}, \bibinfo
  {author} {\bibfnamefont {X.}~\bibnamefont {Liu}}, \bibinfo {author}
  {\bibfnamefont {X.}~\bibnamefont {Zhang}}, \bibinfo {author} {\bibfnamefont
  {Y.}~\bibnamefont {Ma}}, \bibinfo {author} {\bibfnamefont {S.}~\bibnamefont
  {Zhang}}, \bibinfo {author} {\bibfnamefont {S.~A.}\ \bibnamefont {Maier}},
  \bibinfo {author} {\bibfnamefont {Z.}~\bibnamefont {Tian}}, \bibinfo {author}
  {\bibfnamefont {A.~K.}\ \bibnamefont {Azad}}, \bibinfo {author}
  {\bibfnamefont {H.-T.}\ \bibnamefont {Chen}}, \emph {et~al.},\ }\bibfield
  {title} {\bibinfo {title} {Active control of electromagnetically induced
  transparency analogue in terahertz metamaterials},\ }\href
  {https://doi.org/10.1038/ncomms2153} {\bibfield  {journal} {\bibinfo
  {journal} {Nat. Commun.}\ }\textbf {\bibinfo {volume} {3}},\ \bibinfo {pages}
  {1} (\bibinfo {year} {2012})}\BibitemShut {NoStop}%
\bibitem [{\citenamefont {Ma}\ and\ \citenamefont
  {Oulton}(2019)}]{Ma_NatureNano_2019}%
  \BibitemOpen
  \bibfield  {author} {\bibinfo {author} {\bibfnamefont {R.-M.}\ \bibnamefont
  {Ma}}\ and\ \bibinfo {author} {\bibfnamefont {R.~F.}\ \bibnamefont
  {Oulton}},\ }\bibfield  {title} {\bibinfo {title} {Applications of
  nanolasers},\ }\href
  {https://doi.org/https://doi.org/10.1038/s41565-018-0320-y} {\bibfield
  {journal} {\bibinfo  {journal} {Nat. Nanotechnol.}\ }\textbf {\bibinfo
  {volume} {14}},\ \bibinfo {pages} {12} (\bibinfo {year} {2019})}\BibitemShut
  {NoStop}%
\bibitem [{\citenamefont {Cui}\ \emph {et~al.}(2019)\citenamefont {Cui},
  \citenamefont {Bai},\ and\ \citenamefont {Sun}}]{Tong_AdvFuncMat_2019}%
  \BibitemOpen
  \bibfield  {author} {\bibinfo {author} {\bibfnamefont {T.}~\bibnamefont
  {Cui}}, \bibinfo {author} {\bibfnamefont {B.}~\bibnamefont {Bai}},\ and\
  \bibinfo {author} {\bibfnamefont {H.-B.}\ \bibnamefont {Sun}},\ }\bibfield
  {title} {\bibinfo {title} {Tunable metasurfaces based on active materials},\
  }\href {https://doi.org/https://doi.org/10.1002/adfm.201806692} {\bibfield
  {journal} {\bibinfo  {journal} {Adv. Funct. Mater.}\ }\textbf {\bibinfo
  {volume} {29}},\ \bibinfo {pages} {1806692} (\bibinfo {year}
  {2019})}\BibitemShut {NoStop}%
\bibitem [{\citenamefont {Shaltout}\ \emph {et~al.}(2019)\citenamefont
  {Shaltout}, \citenamefont {Shalaev},\ and\ \citenamefont
  {Brongersma}}]{Shaltout_Scince_2019}%
  \BibitemOpen
  \bibfield  {author} {\bibinfo {author} {\bibfnamefont {A.~M.}\ \bibnamefont
  {Shaltout}}, \bibinfo {author} {\bibfnamefont {V.~M.}\ \bibnamefont
  {Shalaev}},\ and\ \bibinfo {author} {\bibfnamefont {M.~L.}\ \bibnamefont
  {Brongersma}},\ }\bibfield  {title} {\bibinfo {title} {Spatiotemporal light
  control with active metasurfaces},\ }\href
  {https://doi.org/10.1126/science.aat3100} {\bibfield  {journal} {\bibinfo
  {journal} {Science}\ }\textbf {\bibinfo {volume} {364}},\ \bibinfo {pages}
  {eaat3100} (\bibinfo {year} {2019})}\BibitemShut {NoStop}%
\bibitem [{\citenamefont {Rahmani}\ \emph {et~al.}(2017)\citenamefont
  {Rahmani}, \citenamefont {Xu}, \citenamefont {Miroshnichenko}, \citenamefont
  {Komar}, \citenamefont {Camacho-Morales}, \citenamefont {Chen}, \citenamefont
  {Z{\'a}rate}, \citenamefont {Kruk}, \citenamefont {Zhang}, \citenamefont
  {Neshev} \emph {et~al.}}]{Rahmani_AdvFuncMat_2017}%
  \BibitemOpen
  \bibfield  {author} {\bibinfo {author} {\bibfnamefont {M.}~\bibnamefont
  {Rahmani}}, \bibinfo {author} {\bibfnamefont {L.}~\bibnamefont {Xu}},
  \bibinfo {author} {\bibfnamefont {A.~E.}\ \bibnamefont {Miroshnichenko}},
  \bibinfo {author} {\bibfnamefont {A.}~\bibnamefont {Komar}}, \bibinfo
  {author} {\bibfnamefont {R.}~\bibnamefont {Camacho-Morales}}, \bibinfo
  {author} {\bibfnamefont {H.}~\bibnamefont {Chen}}, \bibinfo {author}
  {\bibfnamefont {Y.}~\bibnamefont {Z{\'a}rate}}, \bibinfo {author}
  {\bibfnamefont {S.}~\bibnamefont {Kruk}}, \bibinfo {author} {\bibfnamefont
  {G.}~\bibnamefont {Zhang}}, \bibinfo {author} {\bibfnamefont {D.~N.}\
  \bibnamefont {Neshev}}, \emph {et~al.},\ }\bibfield  {title} {\bibinfo
  {title} {Reversible thermal tuning of all-dielectric metasurfaces},\ }\href
  {https://doi.org/https://doi.org/10.1002/adfm.201700580} {\bibfield
  {journal} {\bibinfo  {journal} {Adv. Funct. Mater.}\ }\textbf {\bibinfo
  {volume} {27}},\ \bibinfo {pages} {1700580} (\bibinfo {year}
  {2017})}\BibitemShut {NoStop}%
\bibitem [{\citenamefont {Xiao}\ \emph {et~al.}(2010)\citenamefont {Xiao},
  \citenamefont {Drachev}, \citenamefont {Kildishev}, \citenamefont {Ni},
  \citenamefont {Chettiar}, \citenamefont {Yuan},\ and\ \citenamefont
  {Shalaev}}]{Xiao_Nature_2010}%
  \BibitemOpen
  \bibfield  {author} {\bibinfo {author} {\bibfnamefont {S.}~\bibnamefont
  {Xiao}}, \bibinfo {author} {\bibfnamefont {V.~P.}\ \bibnamefont {Drachev}},
  \bibinfo {author} {\bibfnamefont {A.~V.}\ \bibnamefont {Kildishev}}, \bibinfo
  {author} {\bibfnamefont {X.}~\bibnamefont {Ni}}, \bibinfo {author}
  {\bibfnamefont {U.~K.}\ \bibnamefont {Chettiar}}, \bibinfo {author}
  {\bibfnamefont {H.-K.}\ \bibnamefont {Yuan}},\ and\ \bibinfo {author}
  {\bibfnamefont {V.~M.}\ \bibnamefont {Shalaev}},\ }\bibfield  {title}
  {\bibinfo {title} {Loss-free and active optical negative-index
  metamaterials},\ }\href {https://doi.org/10.1038/nature09278} {\bibfield
  {journal} {\bibinfo  {journal} {Nature}\ }\textbf {\bibinfo {volume} {466}},\
  \bibinfo {pages} {735} (\bibinfo {year} {2010})}\BibitemShut {NoStop}%
\bibitem [{\citenamefont {Amooghorban}\ \emph {et~al.}(2013)\citenamefont
  {Amooghorban}, \citenamefont {Mortensen},\ and\ \citenamefont
  {Wubs}}]{Amooghorban_PRL_2013}%
  \BibitemOpen
  \bibfield  {author} {\bibinfo {author} {\bibfnamefont {E.}~\bibnamefont
  {Amooghorban}}, \bibinfo {author} {\bibfnamefont {N.~A.}\ \bibnamefont
  {Mortensen}},\ and\ \bibinfo {author} {\bibfnamefont {M.}~\bibnamefont
  {Wubs}},\ }\bibfield  {title} {\bibinfo {title} {Quantum optical
  effective-medium theory for loss-compensated metamaterials},\ }\href
  {https://doi.org/10.1103/PhysRevLett.110.153602} {\bibfield  {journal}
  {\bibinfo  {journal} {Phys. Rev. Lett.}\ }\textbf {\bibinfo {volume} {110}},\
  \bibinfo {pages} {153602} (\bibinfo {year} {2013})}\BibitemShut {NoStop}%
\bibitem [{\citenamefont {Ghoshroy}\ \emph {et~al.}(2020)\citenamefont
  {Ghoshroy}, \citenamefont {{\"O}zdemir},\ and\ \citenamefont
  {G{\"u}ney}}]{ghoshroy2020loss}%
  \BibitemOpen
  \bibfield  {author} {\bibinfo {author} {\bibfnamefont {A.}~\bibnamefont
  {Ghoshroy}}, \bibinfo {author} {\bibfnamefont {{\c{S}}.~K.}\ \bibnamefont
  {{\"O}zdemir}},\ and\ \bibinfo {author} {\bibfnamefont {D.~{\"O}.}\
  \bibnamefont {G{\"u}ney}},\ }\bibfield  {title} {\bibinfo {title} {Loss
  compensation in metamaterials and plasmonics with virtual gain},\ }\href
  {https://doi.org/10.1364/OME.397720} {\bibfield  {journal} {\bibinfo
  {journal} {Opt. Mater. Express}\ }\textbf {\bibinfo {volume} {10}},\ \bibinfo
  {pages} {1862} (\bibinfo {year} {2020})}\BibitemShut {NoStop}%
\bibitem [{\citenamefont {Hess}\ \emph {et~al.}(2012)\citenamefont {Hess},
  \citenamefont {Pendry}, \citenamefont {Maier}, \citenamefont {Oulton},
  \citenamefont {Hamm},\ and\ \citenamefont {Tsakmakidis}}]{hess2012active}%
  \BibitemOpen
  \bibfield  {author} {\bibinfo {author} {\bibfnamefont {O.}~\bibnamefont
  {Hess}}, \bibinfo {author} {\bibfnamefont {J.~B.}\ \bibnamefont {Pendry}},
  \bibinfo {author} {\bibfnamefont {S.~A.}\ \bibnamefont {Maier}}, \bibinfo
  {author} {\bibfnamefont {R.~F.}\ \bibnamefont {Oulton}}, \bibinfo {author}
  {\bibfnamefont {J.~M.}\ \bibnamefont {Hamm}},\ and\ \bibinfo {author}
  {\bibfnamefont {K.~L.}\ \bibnamefont {Tsakmakidis}},\ }\bibfield  {title}
  {\bibinfo {title} {Active nanoplasmonic metamaterials},\ }\href
  {https://doi.org/10.1038/nmat3356} {\bibfield  {journal} {\bibinfo  {journal}
  {Nat. Mater.}\ }\textbf {\bibinfo {volume} {11}},\ \bibinfo {pages} {573}
  (\bibinfo {year} {2012})}\BibitemShut {NoStop}%
\bibitem [{\citenamefont {Droulias}\ \emph {et~al.}(2017)\citenamefont
  {Droulias}, \citenamefont {Jain}, \citenamefont {Koschny},\ and\
  \citenamefont {Soukoulis}}]{droulias2017fundamentals}%
  \BibitemOpen
  \bibfield  {author} {\bibinfo {author} {\bibfnamefont {S.}~\bibnamefont
  {Droulias}}, \bibinfo {author} {\bibfnamefont {A.}~\bibnamefont {Jain}},
  \bibinfo {author} {\bibfnamefont {T.}~\bibnamefont {Koschny}},\ and\ \bibinfo
  {author} {\bibfnamefont {C.~M.}\ \bibnamefont {Soukoulis}},\ }\bibfield
  {title} {\bibinfo {title} {Fundamentals of metasurface lasers based on
  resonant dark states},\ }\href {https://doi.org/10.1103/PhysRevB.96.155143}
  {\bibfield  {journal} {\bibinfo  {journal} {Phys. Rev. B}\ }\textbf {\bibinfo
  {volume} {96}},\ \bibinfo {pages} {155143} (\bibinfo {year}
  {2017})}\BibitemShut {NoStop}%
\bibitem [{\citenamefont {Deka}\ \emph {et~al.}(2021)\citenamefont {Deka},
  \citenamefont {Jiang}, \citenamefont {Pan},\ and\ \citenamefont
  {Fainman}}]{deka2021nanolaser}%
  \BibitemOpen
  \bibfield  {author} {\bibinfo {author} {\bibfnamefont {S.~S.}\ \bibnamefont
  {Deka}}, \bibinfo {author} {\bibfnamefont {S.}~\bibnamefont {Jiang}},
  \bibinfo {author} {\bibfnamefont {S.~H.}\ \bibnamefont {Pan}},\ and\ \bibinfo
  {author} {\bibfnamefont {Y.}~\bibnamefont {Fainman}},\ }\bibfield  {title}
  {\bibinfo {title} {Nanolaser arrays: toward application-driven dense
  integration},\ }\href {https://doi.org/doi:10.1515/nanoph-2020-0372}
  {\bibfield  {journal} {\bibinfo  {journal} {Nanophotonics}\ }\textbf
  {\bibinfo {volume} {10}},\ \bibinfo {pages} {149} (\bibinfo {year}
  {2021})}\BibitemShut {NoStop}%
\bibitem [{\citenamefont {Tuz}\ \emph {et~al.}(2010)\citenamefont {Tuz},
  \citenamefont {Prosvirnin},\ and\ \citenamefont
  {Kochetova}}]{tuz_PhysRevB_2010}%
  \BibitemOpen
  \bibfield  {author} {\bibinfo {author} {\bibfnamefont {V.~R.}\ \bibnamefont
  {Tuz}}, \bibinfo {author} {\bibfnamefont {S.~L.}\ \bibnamefont
  {Prosvirnin}},\ and\ \bibinfo {author} {\bibfnamefont {L.~A.}\ \bibnamefont
  {Kochetova}},\ }\bibfield  {title} {\bibinfo {title} {Optical bistability
  involving planar metamaterials with broken structural symmetry},\ }\href
  {https://doi.org/10.1103/PhysRevB.82.233402} {\bibfield  {journal} {\bibinfo
  {journal} {Phys. Rev. B}\ }\textbf {\bibinfo {volume} {82}},\ \bibinfo
  {pages} {233402} (\bibinfo {year} {2010})}\BibitemShut {NoStop}%
\bibitem [{\citenamefont {Tuz}\ \emph {et~al.}(2012)\citenamefont {Tuz},
  \citenamefont {Butylkin},\ and\ \citenamefont {Prosvirnin}}]{tuz_JOpt_2012}%
  \BibitemOpen
  \bibfield  {author} {\bibinfo {author} {\bibfnamefont {V.~R.}\ \bibnamefont
  {Tuz}}, \bibinfo {author} {\bibfnamefont {V.~S.}\ \bibnamefont {Butylkin}},\
  and\ \bibinfo {author} {\bibfnamefont {S.~L.}\ \bibnamefont {Prosvirnin}},\
  }\bibfield  {title} {\bibinfo {title} {Enhancement of absorption bistability
  by trapping-light planar metamaterial},\ }\href
  {https://doi.org/10.1088/2040-8978/14/4/045102} {\bibfield  {journal}
  {\bibinfo  {journal} {J. Opt.}\ }\textbf {\bibinfo {volume} {14}},\ \bibinfo
  {pages} {045102} (\bibinfo {year} {2012})}\BibitemShut {NoStop}%
\bibitem [{\citenamefont {Tuz}\ \emph {et~al.}(2014)\citenamefont {Tuz},
  \citenamefont {Novitsky}, \citenamefont {Mladyonov}, \citenamefont
  {Prosvirnin},\ and\ \citenamefont {Novitsky}}]{Tuz_JOSAB_2014}%
  \BibitemOpen
  \bibfield  {author} {\bibinfo {author} {\bibfnamefont {V.~R.}\ \bibnamefont
  {Tuz}}, \bibinfo {author} {\bibfnamefont {D.~V.}\ \bibnamefont {Novitsky}},
  \bibinfo {author} {\bibfnamefont {P.~L.}\ \bibnamefont {Mladyonov}}, \bibinfo
  {author} {\bibfnamefont {S.~L.}\ \bibnamefont {Prosvirnin}},\ and\ \bibinfo
  {author} {\bibfnamefont {A.~V.}\ \bibnamefont {Novitsky}},\ }\bibfield
  {title} {\bibinfo {title} {Nonlinear interaction of two trapped-mode
  resonances in a bilayer fish-scale metamaterial},\ }\href
  {https://doi.org/10.1364/JOSAB.31.002095} {\bibfield  {journal} {\bibinfo
  {journal} {J. Opt. Soc. Am. B}\ }\textbf {\bibinfo {volume} {31}},\ \bibinfo
  {pages} {2095} (\bibinfo {year} {2014})}\BibitemShut {NoStop}%
\bibitem [{\citenamefont {Dyachenko}\ \emph {et~al.}(2016)\citenamefont
  {Dyachenko}, \citenamefont {Molesky}, \citenamefont {Petrov}, \citenamefont
  {St{\"o}rmer}, \citenamefont {Krekeler}, \citenamefont {Lang}, \citenamefont
  {Ritter}, \citenamefont {Jacob},\ and\ \citenamefont
  {Eich}}]{dyachenko2016controlling}%
  \BibitemOpen
  \bibfield  {author} {\bibinfo {author} {\bibfnamefont {P.~N.}\ \bibnamefont
  {Dyachenko}}, \bibinfo {author} {\bibfnamefont {S.}~\bibnamefont {Molesky}},
  \bibinfo {author} {\bibfnamefont {A.~Y.}\ \bibnamefont {Petrov}}, \bibinfo
  {author} {\bibfnamefont {M.}~\bibnamefont {St{\"o}rmer}}, \bibinfo {author}
  {\bibfnamefont {T.}~\bibnamefont {Krekeler}}, \bibinfo {author}
  {\bibfnamefont {S.}~\bibnamefont {Lang}}, \bibinfo {author} {\bibfnamefont
  {M.}~\bibnamefont {Ritter}}, \bibinfo {author} {\bibfnamefont
  {Z.}~\bibnamefont {Jacob}},\ and\ \bibinfo {author} {\bibfnamefont
  {M.}~\bibnamefont {Eich}},\ }\bibfield  {title} {\bibinfo {title}
  {Controlling thermal emission with refractory epsilon-near-zero metamaterials
  via topological transitions},\ }\href {https://doi.org/10.1038/ncomms11809}
  {\bibfield  {journal} {\bibinfo  {journal} {Nat. Commun.}\ }\textbf {\bibinfo
  {volume} {7}},\ \bibinfo {pages} {1} (\bibinfo {year} {2016})}\BibitemShut
  {NoStop}%
\bibitem [{\citenamefont {Dabidian}\ \emph {et~al.}(2016)\citenamefont
  {Dabidian}, \citenamefont {Dutta-Gupta}, \citenamefont {Kholmanov},
  \citenamefont {Lai}, \citenamefont {Lu}, \citenamefont {Lee}, \citenamefont
  {Jin}, \citenamefont {Trendafilov}, \citenamefont {Khanikaev}, \citenamefont
  {Fallahazad} \emph {et~al.}}]{dabidian2016experimental}%
  \BibitemOpen
  \bibfield  {author} {\bibinfo {author} {\bibfnamefont {N.}~\bibnamefont
  {Dabidian}}, \bibinfo {author} {\bibfnamefont {S.}~\bibnamefont
  {Dutta-Gupta}}, \bibinfo {author} {\bibfnamefont {I.}~\bibnamefont
  {Kholmanov}}, \bibinfo {author} {\bibfnamefont {K.}~\bibnamefont {Lai}},
  \bibinfo {author} {\bibfnamefont {F.}~\bibnamefont {Lu}}, \bibinfo {author}
  {\bibfnamefont {J.}~\bibnamefont {Lee}}, \bibinfo {author} {\bibfnamefont
  {M.}~\bibnamefont {Jin}}, \bibinfo {author} {\bibfnamefont {S.}~\bibnamefont
  {Trendafilov}}, \bibinfo {author} {\bibfnamefont {A.}~\bibnamefont
  {Khanikaev}}, \bibinfo {author} {\bibfnamefont {B.}~\bibnamefont
  {Fallahazad}}, \emph {et~al.},\ }\bibfield  {title} {\bibinfo {title}
  {Experimental demonstration of phase modulation and motion sensing using
  graphene-integrated metasurfaces},\ }\href
  {https://doi.org/10.1021/acs.nanolett.6b00732} {\bibfield  {journal}
  {\bibinfo  {journal} {Nano Lett.}\ }\textbf {\bibinfo {volume} {16}},\
  \bibinfo {pages} {3607} (\bibinfo {year} {2016})}\BibitemShut {NoStop}%
\bibitem [{\citenamefont {Li}\ \emph {et~al.}(2017)\citenamefont {Li},
  \citenamefont {Jun~Cui}, \citenamefont {Ji}, \citenamefont {Liu},
  \citenamefont {Ding}, \citenamefont {Wan}, \citenamefont {Bo~Li},
  \citenamefont {Jiang}, \citenamefont {Qiu},\ and\ \citenamefont
  {Zhang}}]{li2017electromagnetic}%
  \BibitemOpen
  \bibfield  {author} {\bibinfo {author} {\bibfnamefont {L.}~\bibnamefont
  {Li}}, \bibinfo {author} {\bibfnamefont {T.}~\bibnamefont {Jun~Cui}},
  \bibinfo {author} {\bibfnamefont {W.}~\bibnamefont {Ji}}, \bibinfo {author}
  {\bibfnamefont {S.}~\bibnamefont {Liu}}, \bibinfo {author} {\bibfnamefont
  {J.}~\bibnamefont {Ding}}, \bibinfo {author} {\bibfnamefont {X.}~\bibnamefont
  {Wan}}, \bibinfo {author} {\bibfnamefont {Y.}~\bibnamefont {Bo~Li}}, \bibinfo
  {author} {\bibfnamefont {M.}~\bibnamefont {Jiang}}, \bibinfo {author}
  {\bibfnamefont {C.-W.}\ \bibnamefont {Qiu}},\ and\ \bibinfo {author}
  {\bibfnamefont {S.}~\bibnamefont {Zhang}},\ }\bibfield  {title} {\bibinfo
  {title} {Electromagnetic reprogrammable coding-metasurface holograms},\
  }\href {https://doi.org/10.1038/s41467-017-00164-9} {\bibfield  {journal}
  {\bibinfo  {journal} {Nat. Commun.}\ }\textbf {\bibinfo {volume} {8}},\
  \bibinfo {pages} {1} (\bibinfo {year} {2017})}\BibitemShut {NoStop}%
\bibitem [{\citenamefont {Malek}\ \emph {et~al.}(2017)\citenamefont {Malek},
  \citenamefont {Ee},\ and\ \citenamefont {Agarwal}}]{malek2017strain}%
  \BibitemOpen
  \bibfield  {author} {\bibinfo {author} {\bibfnamefont {S.~C.}\ \bibnamefont
  {Malek}}, \bibinfo {author} {\bibfnamefont {H.-S.}\ \bibnamefont {Ee}},\ and\
  \bibinfo {author} {\bibfnamefont {R.}~\bibnamefont {Agarwal}},\ }\bibfield
  {title} {\bibinfo {title} {Strain multiplexed metasurface holograms on a
  stretchable substrate},\ }\href
  {https://doi.org/10.1021/acs.nanolett.7b00807} {\bibfield  {journal}
  {\bibinfo  {journal} {Nano Lett.}\ }\textbf {\bibinfo {volume} {17}},\
  \bibinfo {pages} {3641} (\bibinfo {year} {2017})}\BibitemShut {NoStop}%
\bibitem [{\citenamefont {Khardikov}\ \emph {et~al.}(2012)\citenamefont
  {Khardikov}, \citenamefont {Iarko},\ and\ \citenamefont
  {Prosvirnin}}]{khardikov2012giant}%
  \BibitemOpen
  \bibfield  {author} {\bibinfo {author} {\bibfnamefont {V.~V.}\ \bibnamefont
  {Khardikov}}, \bibinfo {author} {\bibfnamefont {E.~O.}\ \bibnamefont
  {Iarko}},\ and\ \bibinfo {author} {\bibfnamefont {S.~L.}\ \bibnamefont
  {Prosvirnin}},\ }\bibfield  {title} {\bibinfo {title} {A giant red shift and
  enhancement of the light confinement in a planar array of dielectric bars},\
  }\href {https://doi.org/10.1088/2040-8978/14/3/035103} {\bibfield  {journal}
  {\bibinfo  {journal} {J. Opt.}\ }\textbf {\bibinfo {volume} {14}},\ \bibinfo
  {pages} {035103} (\bibinfo {year} {2012})}\BibitemShut {NoStop}%
\bibitem [{\citenamefont {Zhang}\ \emph {et~al.}(2013)\citenamefont {Zhang},
  \citenamefont {MacDonald},\ and\ \citenamefont {Zheludev}}]{zhang2013near}%
  \BibitemOpen
  \bibfield  {author} {\bibinfo {author} {\bibfnamefont {J.}~\bibnamefont
  {Zhang}}, \bibinfo {author} {\bibfnamefont {K.~F.}\ \bibnamefont
  {MacDonald}},\ and\ \bibinfo {author} {\bibfnamefont {N.~I.}\ \bibnamefont
  {Zheludev}},\ }\bibfield  {title} {\bibinfo {title} {Near-infrared trapped
  mode magnetic resonance in an all-dielectric metamaterial},\ }\href
  {https://doi.org/10.1364/OE.21.026721} {\bibfield  {journal} {\bibinfo
  {journal} {Opt. Express}\ }\textbf {\bibinfo {volume} {21}},\ \bibinfo
  {pages} {26721} (\bibinfo {year} {2013})}\BibitemShut {NoStop}%
\bibitem [{\citenamefont {Khardikov}\ \emph {et~al.}(2016)\citenamefont
  {Khardikov}, \citenamefont {Mladyonov}, \citenamefont {Prosvirnin},\ and\
  \citenamefont {Tuz}}]{Khardikov2016}%
  \BibitemOpen
  \bibfield  {author} {\bibinfo {author} {\bibfnamefont {V.}~\bibnamefont
  {Khardikov}}, \bibinfo {author} {\bibfnamefont {P.}~\bibnamefont
  {Mladyonov}}, \bibinfo {author} {\bibfnamefont {S.}~\bibnamefont
  {Prosvirnin}},\ and\ \bibinfo {author} {\bibfnamefont {V.}~\bibnamefont
  {Tuz}},\ }\bibinfo {title} {Electromagnetic wave diffraction by periodic
  planar metamaterials with nonlinear constituents},\ in\ \href
  {https://doi.org/10.1007/978-94-017-7315-7_5} {\emph {\bibinfo {booktitle}
  {Contemporary Optoelectronics: Materials, Metamaterials and Device
  Applications}}},\ \bibinfo {editor} {edited by\ \bibinfo {editor}
  {\bibfnamefont {O.}~\bibnamefont {Shulika}}\ and\ \bibinfo {editor}
  {\bibfnamefont {I.}~\bibnamefont {Sukhoivanov}}}\ (\bibinfo  {publisher}
  {Springer Netherlands},\ \bibinfo {address} {Dordrecht},\ \bibinfo {year}
  {2016})\ Chap.~\bibinfo {chapter} {5}, pp.\ \bibinfo {pages}
  {81--98}\BibitemShut {NoStop}%
\bibitem [{\citenamefont {Tuz}\ \emph {et~al.}(2018)\citenamefont {Tuz},
  \citenamefont {Khardikov}, \citenamefont {Kupriianov}, \citenamefont
  {Domina}, \citenamefont {Xu}, \citenamefont {Wang},\ and\ \citenamefont
  {Sun}}]{tuz2018high}%
  \BibitemOpen
  \bibfield  {author} {\bibinfo {author} {\bibfnamefont {V.~R.}\ \bibnamefont
  {Tuz}}, \bibinfo {author} {\bibfnamefont {V.~V.}\ \bibnamefont {Khardikov}},
  \bibinfo {author} {\bibfnamefont {A.~S.}\ \bibnamefont {Kupriianov}},
  \bibinfo {author} {\bibfnamefont {K.~L.}\ \bibnamefont {Domina}}, \bibinfo
  {author} {\bibfnamefont {S.}~\bibnamefont {Xu}}, \bibinfo {author}
  {\bibfnamefont {H.}~\bibnamefont {Wang}},\ and\ \bibinfo {author}
  {\bibfnamefont {H.-B.}\ \bibnamefont {Sun}},\ }\bibfield  {title} {\bibinfo
  {title} {High-quality trapped modes in all-dielectric metamaterials},\ }\href
  {https://doi.org/10.1364/OE.26.002905} {\bibfield  {journal} {\bibinfo
  {journal} {Opt. Express}\ }\textbf {\bibinfo {volume} {26}},\ \bibinfo
  {pages} {2905} (\bibinfo {year} {2018})}\BibitemShut {NoStop}%
\bibitem [{\citenamefont {Hsu}\ \emph {et~al.}(2016)\citenamefont {Hsu},
  \citenamefont {Zhen}, \citenamefont {Stone}, \citenamefont {Joannopoulos},\
  and\ \citenamefont {Soljačić}}]{Hsu2016BIC}%
  \BibitemOpen
  \bibfield  {author} {\bibinfo {author} {\bibfnamefont {C.}~\bibnamefont
  {Hsu}}, \bibinfo {author} {\bibfnamefont {B.}~\bibnamefont {Zhen}}, \bibinfo
  {author} {\bibfnamefont {A.~D.}\ \bibnamefont {Stone}}, \bibinfo {author}
  {\bibfnamefont {J.~D.}\ \bibnamefont {Joannopoulos}},\ and\ \bibinfo {author}
  {\bibfnamefont {M.}~\bibnamefont {Soljačić}},\ }\bibfield  {title}
  {\bibinfo {title} {Bound states in the continuum},\ }\href
  {https://doi.org/https://doi.org/10.1038/natrevmats.2016.48} {\bibfield
  {journal} {\bibinfo  {journal} {Nat. Rev. Mater.}\ }\textbf {\bibinfo
  {volume} {1}},\ \bibinfo {pages} {16048} (\bibinfo {year}
  {2016})}\BibitemShut {NoStop}%
\bibitem [{\citenamefont {Azzam}\ and\ \citenamefont
  {Kildishev}(2021)}]{Azzam2020AOM}%
  \BibitemOpen
  \bibfield  {author} {\bibinfo {author} {\bibfnamefont {S.~I.}\ \bibnamefont
  {Azzam}}\ and\ \bibinfo {author} {\bibfnamefont {A.~V.}\ \bibnamefont
  {Kildishev}},\ }\bibfield  {title} {\bibinfo {title} {Photonic bound states
  in the continuum: From basics to applications},\ }\href
  {https://doi.org/https://doi.org/10.1002/adom.202001469} {\bibfield
  {journal} {\bibinfo  {journal} {Adv. Opt. Mater.}\ }\textbf {\bibinfo
  {volume} {9}},\ \bibinfo {pages} {2001469} (\bibinfo {year}
  {2021})}\BibitemShut {NoStop}%
\bibitem [{\citenamefont {Sadreev}(2021)}]{Sadreev2021rpp}%
  \BibitemOpen
  \bibfield  {author} {\bibinfo {author} {\bibfnamefont {A.~F.}\ \bibnamefont
  {Sadreev}},\ }\bibfield  {title} {\bibinfo {title} {Interference traps waves
  in an open system: bound states in the continuum},\ }\href
  {https://doi.org/https://doi.org/10.1088/1361-6633/abefb9} {\bibfield
  {journal} {\bibinfo  {journal} {Rep. Prog. Phys.}\ }\textbf {\bibinfo
  {volume} {84}},\ \bibinfo {pages} {055901} (\bibinfo {year}
  {2021})}\BibitemShut {NoStop}%
\bibitem [{\citenamefont {Fedotov}\ \emph {et~al.}(2007)\citenamefont
  {Fedotov}, \citenamefont {Rose}, \citenamefont {Prosvirnin}, \citenamefont
  {Papasimakis},\ and\ \citenamefont {Zheludev}}]{fedotov2007sharp}%
  \BibitemOpen
  \bibfield  {author} {\bibinfo {author} {\bibfnamefont {V.~A.}\ \bibnamefont
  {Fedotov}}, \bibinfo {author} {\bibfnamefont {M.}~\bibnamefont {Rose}},
  \bibinfo {author} {\bibfnamefont {S.~L.}\ \bibnamefont {Prosvirnin}},
  \bibinfo {author} {\bibfnamefont {N.}~\bibnamefont {Papasimakis}},\ and\
  \bibinfo {author} {\bibfnamefont {N.~I.}\ \bibnamefont {Zheludev}},\
  }\bibfield  {title} {\bibinfo {title} {Sharp trapped-mode resonances in
  planar metamaterials with a broken structural symmetry},\ }\href
  {https://doi.org/10.1103/PhysRevLett.99.147401} {\bibfield  {journal}
  {\bibinfo  {journal} {Phys. Rev. Lett.}\ }\textbf {\bibinfo {volume} {99}},\
  \bibinfo {pages} {147401} (\bibinfo {year} {2007})}\BibitemShut {NoStop}%
\bibitem [{\citenamefont {Tian}\ \emph {et~al.}(2014)\citenamefont {Tian},
  \citenamefont {Fang},\ and\ \citenamefont {Zhang}}]{Tian_ACSPhoton_2014}%
  \BibitemOpen
  \bibfield  {author} {\bibinfo {author} {\bibfnamefont {X.}~\bibnamefont
  {Tian}}, \bibinfo {author} {\bibfnamefont {Y.}~\bibnamefont {Fang}},\ and\
  \bibinfo {author} {\bibfnamefont {B.}~\bibnamefont {Zhang}},\ }\bibfield
  {title} {\bibinfo {title} {Multipolar {Fano} resonances and {Fano}-assisted
  optical activity in silver nanorice heterodimers},\ }\href
  {https://doi.org/10.1021/ph5002457} {\bibfield  {journal} {\bibinfo
  {journal} {ACS Photonics}\ }\textbf {\bibinfo {volume} {1}},\ \bibinfo
  {pages} {1156} (\bibinfo {year} {2014})}\BibitemShut {NoStop}%
\bibitem [{\citenamefont {Fan}\ \emph {et~al.}(2019)\citenamefont {Fan},
  \citenamefont {Shadrivov},\ and\ \citenamefont {Padilla}}]{Fan_Optica_2019}%
  \BibitemOpen
  \bibfield  {author} {\bibinfo {author} {\bibfnamefont {K.}~\bibnamefont
  {Fan}}, \bibinfo {author} {\bibfnamefont {I.~V.}\ \bibnamefont {Shadrivov}},\
  and\ \bibinfo {author} {\bibfnamefont {W.~J.}\ \bibnamefont {Padilla}},\
  }\bibfield  {title} {\bibinfo {title} {Dynamic bound states in the
  continuum},\ }\href {https://doi.org/10.1364/OPTICA.6.000169} {\bibfield
  {journal} {\bibinfo  {journal} {Optica}\ }\textbf {\bibinfo {volume} {6}},\
  \bibinfo {pages} {169} (\bibinfo {year} {2019})}\BibitemShut {NoStop}%
\bibitem [{\citenamefont {van Hoof}\ \emph {et~al.}(2021)\citenamefont {van
  Hoof}, \citenamefont {Abujetas}, \citenamefont {Ter~Huurne}, \citenamefont
  {Verdelli}, \citenamefont {Timmermans}, \citenamefont {S{\'a}nchez-Gil},\
  and\ \citenamefont {Rivas}}]{van2021unveiling}%
  \BibitemOpen
  \bibfield  {author} {\bibinfo {author} {\bibfnamefont {N.~J.~J.}\
  \bibnamefont {van Hoof}}, \bibinfo {author} {\bibfnamefont {D.~R.}\
  \bibnamefont {Abujetas}}, \bibinfo {author} {\bibfnamefont {S.~E.~T.}\
  \bibnamefont {Ter~Huurne}}, \bibinfo {author} {\bibfnamefont
  {F.}~\bibnamefont {Verdelli}}, \bibinfo {author} {\bibfnamefont {G.~C.~A.}\
  \bibnamefont {Timmermans}}, \bibinfo {author} {\bibfnamefont {J.~A.}\
  \bibnamefont {S{\'a}nchez-Gil}},\ and\ \bibinfo {author} {\bibfnamefont
  {J.~G.}\ \bibnamefont {Rivas}},\ }\bibfield  {title} {\bibinfo {title}
  {Unveiling the symmetry protection of bound states in the continuum with
  terahertz near-field imaging},\ }\href
  {https://doi.org/10.1021/acsphotonics.1c00937} {\bibfield  {journal}
  {\bibinfo  {journal} {ACS Photonics}\ }\textbf {\bibinfo {volume} {8}},\
  \bibinfo {pages} {3010} (\bibinfo {year} {2021})}\BibitemShut {NoStop}%
\bibitem [{\citenamefont {Ndao}\ \emph {et~al.}(2020)\citenamefont {Ndao},
  \citenamefont {Hsu}, \citenamefont {Cai}, \citenamefont {Ha}, \citenamefont
  {Park}, \citenamefont {Contractor}, \citenamefont {Lo},\ and\ \citenamefont
  {Kant{\'e}}}]{Ndao_Nanophotonics_2020}%
  \BibitemOpen
  \bibfield  {author} {\bibinfo {author} {\bibfnamefont {A.}~\bibnamefont
  {Ndao}}, \bibinfo {author} {\bibfnamefont {L.}~\bibnamefont {Hsu}}, \bibinfo
  {author} {\bibfnamefont {W.}~\bibnamefont {Cai}}, \bibinfo {author}
  {\bibfnamefont {J.}~\bibnamefont {Ha}}, \bibinfo {author} {\bibfnamefont
  {J.}~\bibnamefont {Park}}, \bibinfo {author} {\bibfnamefont {R.}~\bibnamefont
  {Contractor}}, \bibinfo {author} {\bibfnamefont {Y.}~\bibnamefont {Lo}},\
  and\ \bibinfo {author} {\bibfnamefont {B.}~\bibnamefont {Kant{\'e}}},\
  }\bibfield  {title} {\bibinfo {title} {Differentiating and quantifying
  exosome secretion from a single cell using quasi-bound states in the
  continuum},\ }\href {https://doi.org/doi:10.1515/nanoph-2020-0008} {\bibfield
   {journal} {\bibinfo  {journal} {Nanophotonics}\ }\textbf {\bibinfo {volume}
  {9}},\ \bibinfo {pages} {1081} (\bibinfo {year} {2020})}\BibitemShut
  {NoStop}%
\bibitem [{\citenamefont {Gorkunov}\ \emph {et~al.}(2021)\citenamefont
  {Gorkunov}, \citenamefont {Antonov}, \citenamefont {Tuz}, \citenamefont
  {Kupriianov},\ and\ \citenamefont {Kivshar}}]{gorkunov2021bound}%
  \BibitemOpen
  \bibfield  {author} {\bibinfo {author} {\bibfnamefont {M.~V.}\ \bibnamefont
  {Gorkunov}}, \bibinfo {author} {\bibfnamefont {A.~A.}\ \bibnamefont
  {Antonov}}, \bibinfo {author} {\bibfnamefont {V.~R.}\ \bibnamefont {Tuz}},
  \bibinfo {author} {\bibfnamefont {A.~S.}\ \bibnamefont {Kupriianov}},\ and\
  \bibinfo {author} {\bibfnamefont {Y.~S.}\ \bibnamefont {Kivshar}},\
  }\bibfield  {title} {\bibinfo {title} {Bound states in the continuum underpin
  near-lossless maximum chirality in dielectric metasurfaces},\ }\href
  {https://doi.org/https://doi.org/10.1002/adom.202100797} {\bibfield
  {journal} {\bibinfo  {journal} {Adv. Opt. Mater.}\ }\textbf {\bibinfo
  {volume} {9}},\ \bibinfo {pages} {2100797} (\bibinfo {year}
  {2021})}\BibitemShut {NoStop}%
\bibitem [{\citenamefont {Kuznetsov}\ \emph {et~al.}(2021)\citenamefont
  {Kuznetsov}, \citenamefont {Valero}, \citenamefont {Tarkhov}, \citenamefont
  {Bobrovs}, \citenamefont {Redka},\ and\ \citenamefont
  {Shalin}}]{kuznetsov2021transparent}%
  \BibitemOpen
  \bibfield  {author} {\bibinfo {author} {\bibfnamefont {A.~V.}\ \bibnamefont
  {Kuznetsov}}, \bibinfo {author} {\bibfnamefont {A.~C.}\ \bibnamefont
  {Valero}}, \bibinfo {author} {\bibfnamefont {M.}~\bibnamefont {Tarkhov}},
  \bibinfo {author} {\bibfnamefont {V.}~\bibnamefont {Bobrovs}}, \bibinfo
  {author} {\bibfnamefont {D.}~\bibnamefont {Redka}},\ and\ \bibinfo {author}
  {\bibfnamefont {A.~S.}\ \bibnamefont {Shalin}},\ }\bibfield  {title}
  {\bibinfo {title} {Transparent hybrid anapole metasurfaces with negligible
  electromagnetic coupling for phase engineering},\ }\href
  {https://doi.org/doi:10.1515/nanoph-2021-0377} {\bibfield  {journal}
  {\bibinfo  {journal} {Nanophotonics}\ }\textbf {\bibinfo {volume} {10}},\
  \bibinfo {pages} {4385} (\bibinfo {year} {2021})}\BibitemShut {NoStop}%
\bibitem [{\citenamefont {Novitsky}\ \emph {et~al.}(2021)\citenamefont
  {Novitsky}, \citenamefont {Shalin}, \citenamefont {Redka}, \citenamefont
  {Bobrovs},\ and\ \citenamefont {Novitsky}}]{Novitsky2021PRB}%
  \BibitemOpen
  \bibfield  {author} {\bibinfo {author} {\bibfnamefont {D.~V.}\ \bibnamefont
  {Novitsky}}, \bibinfo {author} {\bibfnamefont {A.~S.}\ \bibnamefont
  {Shalin}}, \bibinfo {author} {\bibfnamefont {D.}~\bibnamefont {Redka}},
  \bibinfo {author} {\bibfnamefont {V.}~\bibnamefont {Bobrovs}},\ and\ \bibinfo
  {author} {\bibfnamefont {A.~V.}\ \bibnamefont {Novitsky}},\ }\bibfield
  {title} {\bibinfo {title} {Quasibound states in the continuum induced by
  $\mathcal{PT}$ symmetry breaking},\ }\href
  {https://doi.org/10.1103/PhysRevB.104.085126} {\bibfield  {journal} {\bibinfo
   {journal} {Phys. Rev. B}\ }\textbf {\bibinfo {volume} {104}},\ \bibinfo
  {pages} {085126} (\bibinfo {year} {2021})}\BibitemShut {NoStop}%
\bibitem [{\citenamefont {Weber}(2018)}]{weber2018handbook}%
  \BibitemOpen
  \bibfield  {author} {\bibinfo {author} {\bibfnamefont {M.~J.}\ \bibnamefont
  {Weber}},\ }\href {https://doi.org/10.1201/9781315219639} {\emph {\bibinfo
  {title} {Handbook of Laser Wavelengths}}}\ (\bibinfo  {publisher} {CRC
  press},\ \bibinfo {year} {2018})\BibitemShut {NoStop}%
\bibitem [{\citenamefont {Koshelev}\ \emph {et~al.}(2018)\citenamefont
  {Koshelev}, \citenamefont {Lepeshov}, \citenamefont {Liu}, \citenamefont
  {Bogdanov},\ and\ \citenamefont {Kivshar}}]{Koshelev2018PRL}%
  \BibitemOpen
  \bibfield  {author} {\bibinfo {author} {\bibfnamefont {K.}~\bibnamefont
  {Koshelev}}, \bibinfo {author} {\bibfnamefont {S.}~\bibnamefont {Lepeshov}},
  \bibinfo {author} {\bibfnamefont {M.}~\bibnamefont {Liu}}, \bibinfo {author}
  {\bibfnamefont {A.}~\bibnamefont {Bogdanov}},\ and\ \bibinfo {author}
  {\bibfnamefont {Y.}~\bibnamefont {Kivshar}},\ }\bibfield  {title} {\bibinfo
  {title} {Asymmetric metasurfaces with high-{$Q$} resonances governed by bound
  states in the continuum},\ }\href
  {https://doi.org/10.1103/PhysRevLett.121.193903} {\bibfield  {journal}
  {\bibinfo  {journal} {Phys. Rev. Lett.}\ }\textbf {\bibinfo {volume} {121}},\
  \bibinfo {pages} {193903} (\bibinfo {year} {2018})}\BibitemShut {NoStop}%
\bibitem [{\citenamefont {Xie}\ \emph {et~al.}(1991)\citenamefont {Xie},
  \citenamefont {Fitzgerald},\ and\ \citenamefont {Mii}}]{Xie_JApplPhys_1991}%
  \BibitemOpen
  \bibfield  {author} {\bibinfo {author} {\bibfnamefont {Y.~H.}\ \bibnamefont
  {Xie}}, \bibinfo {author} {\bibfnamefont {E.~A.}\ \bibnamefont
  {Fitzgerald}},\ and\ \bibinfo {author} {\bibfnamefont {Y.~J.}\ \bibnamefont
  {Mii}},\ }\bibfield  {title} {\bibinfo {title} {Evaluation of erbium‐doped
  silicon for optoelectronic applications},\ }\href
  {https://doi.org/10.1063/1.349306} {\bibfield  {journal} {\bibinfo  {journal}
  {J. Appl. Phys.}\ }\textbf {\bibinfo {volume} {70}},\ \bibinfo {pages} {3223}
  (\bibinfo {year} {1991})}\BibitemShut {NoStop}%
\bibitem [{\citenamefont {Thao}\ \emph {et~al.}(2000)\citenamefont {Thao},
  \citenamefont {Ammerlaan},\ and\ \citenamefont
  {Gregorkiewicz}}]{Thao_JApplPhys_2000}%
  \BibitemOpen
  \bibfield  {author} {\bibinfo {author} {\bibfnamefont {D.~T.~X.}\
  \bibnamefont {Thao}}, \bibinfo {author} {\bibfnamefont {C.~A.~J.}\
  \bibnamefont {Ammerlaan}},\ and\ \bibinfo {author} {\bibfnamefont
  {T.}~\bibnamefont {Gregorkiewicz}},\ }\bibfield  {title} {\bibinfo {title}
  {Photoluminescence of erbium-doped silicon: Excitation power and temperature
  dependence},\ }\href {https://doi.org/10.1063/1.373837} {\bibfield  {journal}
  {\bibinfo  {journal} {J. Appl. Phys.}\ }\textbf {\bibinfo {volume} {88}},\
  \bibinfo {pages} {1443} (\bibinfo {year} {2000})}\BibitemShut {NoStop}%
\bibitem [{\citenamefont {Krupka}\ \emph {et~al.}(2005)\citenamefont {Krupka},
  \citenamefont {Tobar}, \citenamefont {Hartnett}, \citenamefont {Cros},\ and\
  \citenamefont {Le~Floch}}]{krupka2005extremely}%
  \BibitemOpen
  \bibfield  {author} {\bibinfo {author} {\bibfnamefont {J.}~\bibnamefont
  {Krupka}}, \bibinfo {author} {\bibfnamefont {M.~E.}\ \bibnamefont {Tobar}},
  \bibinfo {author} {\bibfnamefont {J.~G.}\ \bibnamefont {Hartnett}}, \bibinfo
  {author} {\bibfnamefont {D.}~\bibnamefont {Cros}},\ and\ \bibinfo {author}
  {\bibfnamefont {J.-M.}\ \bibnamefont {Le~Floch}},\ }\bibfield  {title}
  {\bibinfo {title} {Extremely high-{Q} factor dielectric resonators for
  millimeter-wave applications},\ }\href
  {https://doi.org/10.1109/TMTT.2004.840572} {\bibfield  {journal} {\bibinfo
  {journal} {IEEE Trans. Microwave Theory Tech.}\ }\textbf {\bibinfo {volume}
  {53}},\ \bibinfo {pages} {702} (\bibinfo {year} {2005})}\BibitemShut
  {NoStop}%
\bibitem [{\citenamefont {Shcherbinin}\ \emph {et~al.}(2021)\citenamefont
  {Shcherbinin}, \citenamefont {Avramidis}, \citenamefont {Pagonakis},
  \citenamefont {Thumm},\ and\ \citenamefont {Jelonnek}}]{shcherbinin2021IMTW}%
  \BibitemOpen
  \bibfield  {author} {\bibinfo {author} {\bibfnamefont {V.~I.}\ \bibnamefont
  {Shcherbinin}}, \bibinfo {author} {\bibfnamefont {K.~A.}\ \bibnamefont
  {Avramidis}}, \bibinfo {author} {\bibfnamefont {I.~G.}\ \bibnamefont
  {Pagonakis}}, \bibinfo {author} {\bibfnamefont {M.}~\bibnamefont {Thumm}},\
  and\ \bibinfo {author} {\bibfnamefont {J.}~\bibnamefont {Jelonnek}},\
  }\bibfield  {title} {\bibinfo {title} {Large power increase enabled by
  high-{Q} diamond-loaded cavities for terahertz gyrotrons},\ }\href
  {https://doi.org/10.1007/s10762-021-00814-6} {\bibfield  {journal} {\bibinfo
  {journal} {J. Infrared Milli. Terahz. Waves}\ }\textbf {\bibinfo {volume}
  {42}},\ \bibinfo {pages} {863} (\bibinfo {year} {2021})}\BibitemShut
  {NoStop}%
\bibitem [{\citenamefont {Peng}\ \emph {et~al.}(2014)\citenamefont {Peng},
  \citenamefont {{\"O}zdemir}, \citenamefont {Rotter}, \citenamefont {Yilmaz},
  \citenamefont {Liertzer}, \citenamefont {Monifi}, \citenamefont {Bender},
  \citenamefont {Nori},\ and\ \citenamefont {Yang}}]{peng2014loss}%
  \BibitemOpen
  \bibfield  {author} {\bibinfo {author} {\bibfnamefont {B.}~\bibnamefont
  {Peng}}, \bibinfo {author} {\bibfnamefont {{\c{S}}.}~\bibnamefont
  {{\"O}zdemir}}, \bibinfo {author} {\bibfnamefont {S.}~\bibnamefont {Rotter}},
  \bibinfo {author} {\bibfnamefont {H.}~\bibnamefont {Yilmaz}}, \bibinfo
  {author} {\bibfnamefont {M.}~\bibnamefont {Liertzer}}, \bibinfo {author}
  {\bibfnamefont {F.}~\bibnamefont {Monifi}}, \bibinfo {author} {\bibfnamefont
  {C.}~\bibnamefont {Bender}}, \bibinfo {author} {\bibfnamefont
  {F.}~\bibnamefont {Nori}},\ and\ \bibinfo {author} {\bibfnamefont
  {L.}~\bibnamefont {Yang}},\ }\bibfield  {title} {\bibinfo {title}
  {Loss-induced suppression and revival of lasing},\ }\href
  {https://www.science.org/doi/abs/10.1126/science.1258004} {\bibfield
  {journal} {\bibinfo  {journal} {Science}\ }\textbf {\bibinfo {volume}
  {346}},\ \bibinfo {pages} {328} (\bibinfo {year} {2014})}\BibitemShut
  {NoStop}%
\bibitem [{\citenamefont {Sugioka}(2017)}]{sugioka2017progress}%
  \BibitemOpen
  \bibfield  {author} {\bibinfo {author} {\bibfnamefont {K.}~\bibnamefont
  {Sugioka}},\ }\bibfield  {title} {\bibinfo {title} {Progress in ultrafast
  laser processing and future prospects},\ }\href
  {https://doi.org/10.1515/nanoph-2016-0004} {\bibfield  {journal} {\bibinfo
  {journal} {Nanophotonics}\ }\textbf {\bibinfo {volume} {6}},\ \bibinfo
  {pages} {393} (\bibinfo {year} {2017})}\BibitemShut {NoStop}%
\end{thebibliography}%

\end{document}